\documentclass{article}
   
\usepackage[margin=3.5cm, nohead]{geometry} % smaller margins
\usepackage[utf8]{inputenc} % allow utf-8 input
\usepackage[T1]{fontenc}    % use 8-bit T1 fonts
\usepackage{hyperref, enumerate}       % hyperlinks
\usepackage{url}            % simple URL typesetting
\usepackage{booktabs}       % professional-quality tables
\usepackage{amsfonts}       % blackboard math symbols
\usepackage{nicefrac}       % compact symbols for 1/2, etc.
\usepackage{microtype}      % microtypography
\usepackage{xcolor}         % colors
\usepackage{xspace}         % compact space after abbreviations
\usepackage{comment}        % comments
\usepackage{graphicx}       % the [demo] option produces black squares

\usepackage{amssymb,amsmath,dsfont,amsthm}
\newtheorem{definition}{Definition}
\newtheorem{theorem}[definition]{Theorem}

\newtheorem{conjecture}[definition]{Conjecture}
\newtheorem{lemma}[definition]{Lemma}

\newcommand{\proba}{\mathds{P}}
\newcommand{\mean}{\mathds{E}}
\newcommand{\bigO}{\mathcal{O}}
\newcommand{\exactbigO}{\Theta}
\newcommand{\smallo}{o}
\newcommand{\query}{\operatorname{query}}
\newcommand{\queries}{\operatorname{queries}}
\newcommand{\pgf}{\operatorname{PGF}}
\newcommand{\partitions}{\operatorname{Partitions}}
\newcommand{\poch}{\operatorname{Poch}}
\newcommand{\Li}{\operatorname{Li}}
\newcommand{\probaerror}{\proba(\text{undetected error} \mid Q)}

\newcommand*{\eg}{\textit{e.g.}\@\xspace}
\newcommand*{\ie}{\textit{i.e.}\@\xspace}

\newcommand{\ac}{AC }

\usepackage[
]{biblatex}

\title{Active clustering for labeling training data}

\author{
  Quentin Lutz\thanks{Authors presented in alphabetical order.}\\
  Nokia Bell Labs\\
  \and
  \'{E}lie de Panafieu\\
  Nokia Bell Labs\\
  \and
  Alex Scott\\
  University of Oxford\\
  \and
  Maya Stein\\
  University of Chile\\
}

\begin{document}

\maketitle

\begin{abstract}
Gathering training data is a key step of any supervised learning task,
and it is both critical and expensive.
Critical, because the quantity and quality of the training data
has a high impact on the performance of the learned function.
Expensive, because most practical cases rely on humans-in-the-loop to label the data.
The process of determining the correct labels is much more expensive
than comparing two items to see whether they belong to the same class.
Thus motivated, we propose a setting for training data gathering
where the human experts perform the comparatively cheap task of answering pairwise queries,
and the computer groups the items into classes
(which can be labeled cheaply at the very end of the process).
%Unlike in the semi-supervised clustering setting, we do not assume the existence of a distance between the items, or any other additional information.
% Neurips guideline mention that the abstract is only one paragraph.
Given the items, we consider two random models for the classes:
one where the set partition they form is drawn uniformly,
the other one where each item chooses its class independently following a fixed distribution.
In the first model, we characterize the algorithms that
minimize the average number of queries required to cluster the items
and analyze their complexity.
In the second model, we analyze a specific algorithm family,
propose as a conjecture that they reach the minimum average number of queries
and compare their performance to a random approach.
%In the case where all classes have the same size,
%we characterize the algorithms that minimize the average number of queries
%required to cluster the items, and analyze the complexity of these algorithms.
%In the case that classes to have different sizes, the analysis is harder,
%but we propose a conjecture on which algorithms might minimize 
%he average number of queries, and present some evidence for it.
%We also propose solutions to handle errors or inconsistencies in the experts' %answers.
We also propose solutions to handle errors or inconsistencies in the experts' answers.
\end{abstract}

%%%%%%%%%%%%%%%%%%%%%%%%%%%%%%%%%%%%%%%%%%%%%%%%%%%%%%%%%%%
    \section{Introduction}
%%%%%%%%%%%%%%%%%%%%%%%%%%%%%%%%%%%%%%%%%%%%%%%%%%%%%%%%%%%

There is an increasing demand for software implementing
supervised learning for classification.
%Software implementing supervised learning for classification is rapidly growing in quantity and quality.
Training data input for such software consists of items
belonging to distinct classes.
The output is a classifier: a function that predicts,
for any new item, the class it most likely belongs to.
Its quality depends critically
on the available learning data, in terms of both
quantity and quality~\cite{northcutt2021pervasive}.
But labeling large quantities of data is costly.
This task cannot be fully automated, as doing so
would assume access to an already trained classifier.
Thus, human intervention, although expensive, is required.
In this article, we focus on helping the human experts
build the learning data efficiently.

One natural way for the human experts to proceed is to learn (or discover)
the classes and write down their characteristics.
Then, items are considered one by one, assigning an existing class
to each of them, or creating a new one if necessary.
This approach requires the experts to learn the various classes, which,
depending on the use-case, can be difficult.
A different approach, proposed to us by Nokia engineer Maria Laura Maag,
is to discover the partition by querying the experts on two items at a time
asking  whether these belong to the same class or not. 
 This approach avoids the part of the process where classes are learned,
and can therefore be cheaper.
It is the setting we consider here, and we call the corresponding algorithm an {\it active clustering algorithm}, or for short, {\it \ac algorithm}.

More precisely, we assume there is a set of size $n$, with a partition unknown to us. An {\it \ac algorithm} will, in each step, choose a pair of elements and asks the oracle whether they belong to the same partition class or not.
The choices of the queries of the algorithm are allowed to depend on earlier answers.
The algorithm will use transitivity inside the partition classes: if each of the pairs $x$, $y$  and  $y$, $z$ is known to lie in the same class (for instance because of positive answers from the oracle), then the algorithm will `know' that $x$ and $z$ are also in the same class, and it will not submit a query for this pair. The algorithm terminates once the partition is recovered, i.e. when all elements from the same partition class have been shown to belong to the same class, and when for each pair of partition classes, there has been at least one query between their elements. %Two examples are provided in Figure~\ref{fig:ac_examples}

We investigate \ac algorithms under two different random models
for the unknown set partition.
In Section~\ref{sec:uni}, the set partition is sampled uniformly at random,
while in Section~\ref{sec:random},
the number of blocks is fixed and each item chooses its class
independently following the same distribution.
Section~\ref{sec:noisy} analyzes the cases
where the experts' answers contain errors or inconsistencies.
Our proofs, sketched in Section~\ref{sec:proofs},
rely on a broad variety of mathematical tools:
probability theory, graph theory and analytic combinatorics.
We conclude in Section~\ref{sec:conclusion},
with some interesting open problems and research topics.

%\begin{figure}
%\begin{center}
%\caption{Two examples of active clustering starting from the same %situation. Each graph is an aggregated graph (see Section~\ref{sec:uni}),
%the dotted edge is the query considered, the upper (\resp lower) child corresponds to a positive (\resp negative) answer. Assuming uniform distribution on the set partitions, the average complexity are respectively $13/5$ and $12/5$.}
%\label{fig:ac_examples}
%\end{center}
%\end{figure}

%%%%%%%%%%%%
    \paragraph{Related works.}
%%%%%%%%%%%%

Note that our model has similarities with sorting algorithms.
Instead of an array of elements equipped with an order,
we consider a set of items equipped with a set partition structure.
In the sorting algorithm, the %human expert  is an 
oracle determines the 
order of a pair of elements, while in our setting, the oracle tells whether two items
belong to the same class.
Both in sorting and \ac algorithms,
the goal is the design of algorithms 
%with low complexity, where the complexity is defined as the number of queries.
using few queries.

%In \emph{supervised learning}, a function is learned based on training data
%composed of pairs of input, output examples.
%The process of transforming raw data into training data
%is called \emph{annotation}.
%It is a crucial step \cite{northcutt2021pervasive}
%often involving humans-in-the-loop.
The cost of annotating raw data to turn it into training data
motivated the exploration of several variants of supervised learning.
\emph{Transfer learning} reduces the quantity of labeled data required
by using the knowledge gained while solving a different but related problem.
\emph{Semi-supervised learning} reduces it
by learning also from the inherent clustering present in unlabeled data.
In \emph{active learning}, the learner chooses the data needing labeling
and the goal is to maximize the learning potential
for a given small number of queries.
However, users who want to use supervised learning software
for classification as a black-box
can mitigate the annotating cost
only by modifying the labeling process.

Recent papers~\cite{chien2019hs, mazumdar2017clustering} acknowledge that humans prefer pairwise queries over pointwise queries as they are better suited for comparisons.
Pairwise queries have been considered in semi-supervised clustering~\cite{basu2004probabilistic, wagstaff2001constrained, gribel2021semi}
where they are called \emph{must-link} and \emph{cannot-link} constraints.
The pairs of vertices linked by those constraints are random in general,
but chosen adaptively in active semi-supervised clustering~\cite{basu2004active}.
In both cases, the existence of a similarity measure between items
is assumed, and a general theoretical study of this setting is provided by \cite{mazumdar2017sideinfo}.
This is not the case in \cite{eriksson2011active, krishnamurthy2012efficient},
where a hierarchical clustering is built
by successively choosing pairs of items and measuring their similarity.
There, the trade-off between the number of measurements
and the accuracy of the hierarchical clustering is investigated.
The difference with the current paper is that
the similarity measure takes real values,
while we consider Boolean queries,
and that their output is a hierarchical clustering,
while our clustering (a set partition) is flat.

The problem we consider is motivated by its application to annotation
for supervised machine learning.
It also belongs to the field of \emph{combinatorial search} \cite{aigner1988combinatorial, bouvel2005combinatorial}.
Related papers~\cite{alonAsodi,grebinski1999reconstructing}
consider the problem of reconstructing a set partition
using queries on sets of elements,
where the answer to such a query is whether there is an edge in the queried set, or the number of distinct blocks of the partition
present in the queried set, respectively.
Our setting is a more constrained case corresponding to queries of size two.
Another similar setting is that of \emph{entity resolution} \cite{fellegi1969theory} where recent developments also assume perfect oracles and transitive properties using pairwise queries to reconstruct a set partition \cite{vesdapunt2014crowdsourcing, wang2014leveraging}. In this case, clusters correspond to duplicates of the same entity. Most solutions have a real-valued similarity measure between elements but rely on human verification to improve the results.
The entity resolution literature also considers the noisy case for a fixed number of clusters
%Many solutions in the entity resolution literature handle the noisy case for a fixed number of clusters
\cite{cesabianchi2013correlation, chien2019hs, mazumdar2017clustering, davidson2014top}.

\section{Our results}
    \subsection{Uniform distribution on partitions}\label{sec:uni}
%%%%%%%%%%%%%%%%%%%%%%%%%%%%%%%%%%%%%%%%%%%%%%%%%%%%%%%%%%%

We start by considering the setting where the partition of the $n$-set is chosen  uniformly at random among all possible partitions.
%(see the section~\ref{sec:random} for other distributions). 
The \emph{average complexity} of an \ac algorithm
is  the average number of queries used to recover the random partition.
We will define a class of \ac algorithms, called \emph{chordal algorithms},
and prove in Theorem~\ref{th:optimal_chordal}
that an algorithm has minimal average complexity if and only if it is chordal.
Theorem~\ref{th:chordal_same_distribution} shows that all chordal algorithms
have the same distribution on their number of queries.
Finally, this distribution is characterized both exactly and asymptotically
in Theorem~\ref{th:chordal_limit_law}.

\begin{figure}
  \begin{center}
    \includegraphics[width=1\linewidth]{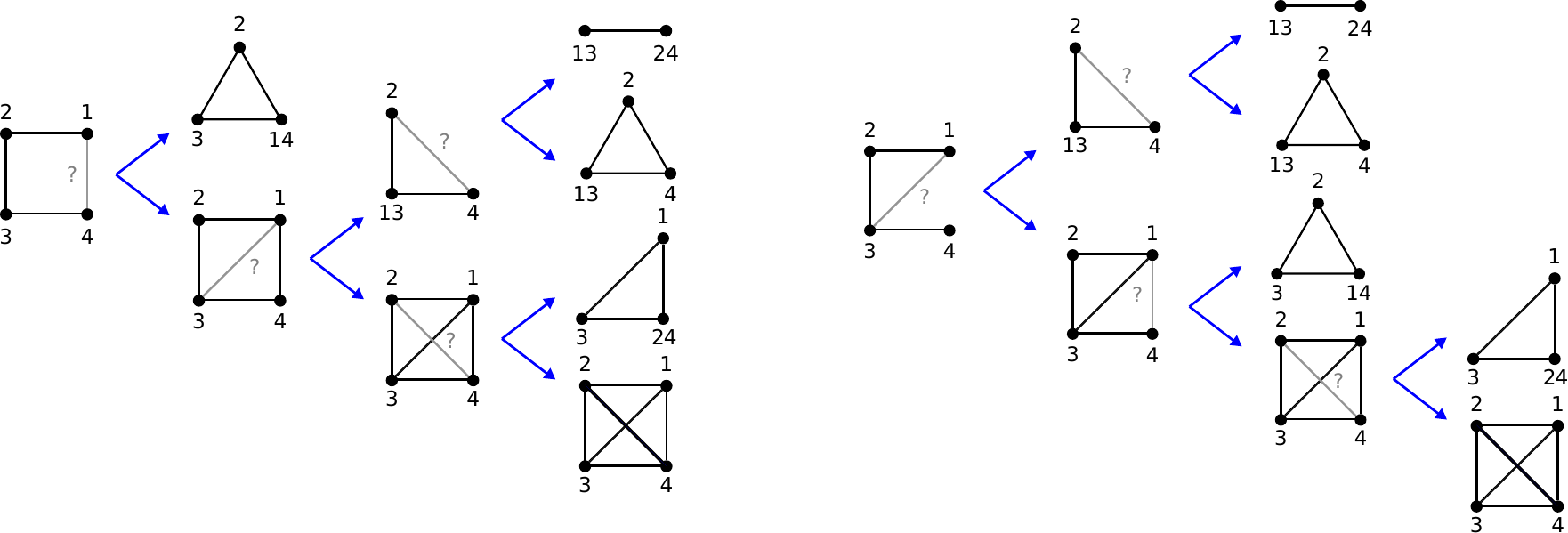}
    \caption{Aggregated graphs obtained after various queries
    and organized into query trees. The positive answers are displayed on top, the negative ones on the bottom. Left, a non-chordal query, leading to an average complexity of $13/5$. Right, chordal queries from the same initial situation, leading to an optimal average complexity of $12/5$.}
  \label{fig:ac_examples}
  \end{center}
\end{figure}

It will be useful to use certain graphs marking the `progress' of the \ac algorithm on our $n$-set.
Given a  partition $P$ of an $n$-set and an \ac algorithm, we can associate a graph $G_t$ to each step $t$ of the algorithm. Namely, $G_0$ has $n$ vertices, labeled with the elements of our set, and at time $t\ge 1$, the graph $G_t$ is obtained by adding an edge for each negatively-answered query, and successively merging pairs of vertices that received a positive answer. 
(A vertex $u$ obtained by joining vertices $v,w$ is adjacent to all neighbors of each of $v,w$, and we label $u$ with the union of its earlier labels.) 
%\marginpar{reworded; and what labels?/A}
We call $G_t$ the {\it aggregated graph} at time $t$. 
Note that  after the last step of the algorithm, the aggregated graph is complete and each of its vertices corresponds to one of the blocks of $P$. Also note that any fixed $G_t$ (with labels on vertices also fixed) may appear as the aggregated graph for more than one partition (possibly at a different time $t'$). We call the set of all these partitions the \emph{realizations} of $G_t$.
Those notions are illustrated in Figure~\ref{fig:ac_examples}.

We now need a quick graph-theoretic definition. A cycle $C$ in a graph is \emph{induced} if all edges induced by $V(C)$ belong to the cycle. % (in other words, $C$ has no \emph{chords}).
A graph is \emph{chordal} if its  induced  cycles all have  length  three. Chordal graphs appear in many applications. 
We say that an \ac algorithm is \emph{chordal} if
one of the following equivalent conditions is satisfied (see the Appendix for a proof of the equivalence of the conditions):
\begin{enumerate}[(i)]
\item\label{chordalA}
for all input partitions, each aggregated graph is chordal,
\item\label{chordalC} for all input partitions, no aggregated graph has an induced $C_4$,
\item\label{chordalB} for all input partitions, and for every query  $u$, $v$ made on some aggregated graph $G_t$, the intersection of the neighborhoods of $u$ and $v$ separates $u$ and $v$ in $G_t$
%is complete and separates $u$ and $v$,
\end{enumerate}
where a set $S$ (possibly empty) is said to \emph{separate} vertices $u,v\in V(G)\setminus S$ if $u$ and $v$ lie in distinct components of $G-S$.
%, and is said to be \emph{complete} if it has all possible edges.
%(Note that in both cases, $S$ is allowed to be empty.)
%
The queries that keep a graph chordal
are exactly those satisfying Condition~\eqref{chordalB}.
Thus, it is used in chordal algorithms
to identify the candidate queries.
On the other hand, a query satisfying Condition~\eqref{chordalC}
might turn a chordal graph non-chordal
by creating an induced cycle of length more than $4$
(in which case, Condition~\eqref{chordalC} will fail on a later query).
%
%Condition~\eqref{chordalB} can be very useful because in order to maintain the aggregated graphs chordal, it is enough to check  at each step of the \ac algorithm that condition \eqref{chordalB} holds. So if in the implementation queries are selected while the algorithm is running, this is a `better' criterion, as it prevents us from having to start anew if we run into a  induced cycle of length larger than 3.
%
Two examples of chordal algorithms are presented below.

\begin{definition} \label{def:clique_universal}
The \emph{clique algorithm} is an \ac algorithm
that gradually grows the partition,
starting with the empty partition.
Each new item is compared successively
to an item of each block of the partition,
until it is either added to a block (positive answer to a query)
or all blocks have been visited,
in which case a new block is created for this item.
The \emph{universal algorithm} finds the block containing the item
of largest label by comparing it to all remaining items.
The algorithm is then applied to partition the other items,
and the previous block is added to the result. 
\end{definition}

\begin{comment}
Figure~\ref{fig:chordal_algorithms}.
\begin{figure}[!htb]
    \begin{minipage}{.5\textwidth}
\begin{verbatim}
def clique_ac(item_set):
    partition = EMPTY_PARTITION
    for u in item_set:
        is_new_block = True
        for block in partition:
            v = an item from block
            if query(u, v):
                block.add(u)
                is_new_block = False
                break
        if is_a_new_block:
            partition.add_block({u})
    return partition
\end{verbatim}
        %\caption{$dt=0.1$}
        %\label{fig:prob1_6_2}
    \end{minipage}%
    \begin{minipage}{0.5\textwidth}
\begin{verbatim}
def universal_ac(item_set):
    if is_empty(item_set):
        return EMPTY_PARTITION
    u = item_set.pop()
    block = {u}
    for v in item_set:
        if query(u, v):
            block.add(v)
            item_set.remove(v)
    partition = universal_ac(item_set)
    partition.add_block(block)
    return partition
\end{verbatim}
        %\caption{$dt =$}
        %\label{fig:prob1_6_1}
    \end{minipage}
    \caption{Two chordal \ac algorithms. The first one is called \emph{universal} because each item is compared to all remaining items. The second one is the \emph{clique algorithm} as after processing each item, the aggregated graph is a clique.}
    \label{fig:chordal_algorithms}
\end{figure}
\end{comment}

%Our first result is the at first sight surprising result that chordal algorithms are the ones that minimize the average  complexity.
\begin{theorem} \label{th:optimal_chordal}
On partitions of size $n$ chosen uniformly at random,
an \ac algorithm has minimal average complexity
if and only if it is chordal.
\end{theorem}

The \emph{complexity distribution} of an \ac algorithm of a set $S$ with $n\ge 1$ elements is a tuple $(a_0, a_1, a_2, \ldots)$ where $a_i$ is the number of partitions of $S$ for which the algorithm has complexity~$i$. Clearly, $a_i=0$ for all $i<n-1$, and $a_{\binom{n}{2}}=1$, for any \ac algorithm.
Our next result shows that constraining the average complexity to be minimal
fixes the complexity distribution.

\begin{theorem} \label{th:chordal_same_distribution}
On partitions of size $n$ chosen uniformly at random,
all chordal \ac algorithms have the same complexity distribution.
\end{theorem}

Note that either of Theorems~\ref{th:optimal_chordal} and~\ref{th:chordal_same_distribution} implies the weaker statement that all chordal \ac algorithms of an $n$-set have the same average complexity, but with very different proofs.
Sketches of the proofs of Theorems~\ref{th:optimal_chordal} and~\ref{th:chordal_same_distribution} can be found in sections~\ref{sec:proof_chordal} and~\ref{sec:sketch_distri}.

In practice, several human experts work in parallel
to annotate the training data.
The chordal algorithms can easily be parallelized:
when an expert becomes available,
give him a query that would keep
the aggregated graph chordal
if all pending queries received negative answers.
The condition ensures the chordality of the parallelized algorithm.

Our third theorem for this setting describes the distribution of the number of queries
used by chordal algorithms on partitions chosen uniformly at random.
Two formulas for the probability generating function are proposed.
The first one is well suited for computer algebra manipulations,
while the second one, more elegant, is better suited for asymptotic analysis.
It is a $q$-analog of Dobi\'nski's formula for Bell numbers
\[
    B_n = \frac{1}{e} \sum_{m \geq 0} \frac{m^n}{n!}
\]
(see \eg \cite[p.762]{FS09}), counting the number of partitions of size $n$.
In order to state the theorem, we introduce the $q$-analogs
of an integer, the factorial, the Pochhammer symbol and the exponential
\[
    [n]_q = \sum_{k=0}^{n-1} q^k,
    \quad
    [n]_q! = \prod_{k=1}^n [k]_q,
    \quad
    (a; q)_n =
    \prod_{k=0}^{n-1}
    (1 - a q^k),
    \quad
    e_q(z) =
    \sum_{n \geq 0}
    \frac{z^n}{[n]_q!}.
\]
Observe that the $q$-analog reduce to their classic counterparts for $q=1$.
The \emph{Lambert function} $W(x)$ is defined for any positive $x$
as the unique solution of the equation $w e^w = x$.
As $x$ tends to infinity, we have $W(x) = \log(x) - \log(\log(x)) + \smallo(1)$.

\begin{theorem} \label{th:chordal_limit_law}
Let $X_n$ denote the complexity of a chordal algorithm
on a partition of size $n$ chosen uniformly at random.
The distribution of $X_n$ has probability generating function
in the variable $q$ equal to the two following expressions
\[
    \frac{1}{B_n}
    \left( \frac{q}{1-q} \right)^n
    \sum_{k = 0}^n
    \binom{n}{k}
    (-1)^k
    \left( \frac{1-q}{q}; q \right)_k
    \qquad \text{and} \qquad
    \frac{1}{B_n}
    \frac{1}{e_q(1/q)}
    \sum_{m \geq 0}
    \frac{[m]_q^n}{[m]_q!}q^{n-m}.
\]
The normalized variable $(X_n - E_n) / \sigma_n$ converges in distribution
to a standard Gaussian law, where
\[
    E_n = \frac{1}{4} (2 W(n) - 1) e^{2 W(n)}
    \qquad \text{and} \qquad
    \sigma_n = \frac{1}{3} \sqrt{\frac{3 W(n)^2 - 4 W(n) + 2}{W(n) + 1} e^{3 W(n)}}.
\]
\end{theorem}

As a corollary, the average complexity of chordal algorithms
is asymptotically $\binom{n}{2} \big/ \log(n)$.
%Consider a data set needing classifying,
%where direct classification by human experts is costly,
%while pairwise queries are cheap.
%If information on the data set is lacking,
%the uniform partition model investigated in this section
%might be pertinent.
Because of this almost quadratic optimal complexity,
using pairwise queries can only be better than direct classification
for relatively small data sets.
We arrive at a different conclusion for the model
presented in the next section,
where \ac algorithm have linear complexity.

%%%%%%%%%%%%%%%%%%%%%%%%%%%%%%%%%%%%%%%%%%%%%%%%%%%%%%%%%%%
    \subsection{Random partitions with fixed number of blocks}\label{sec:random}
%%%%%%%%%%%%%%%%%%%%%%%%%%%%%%%%%%%%%%%%%%%%%%%%%%%%%%%%%%%

In many real-world applications, we have information
on the number of classes and their respective sizes
in the data requiring annotation.
This motivates the introduction of the following
alternative random model for the set partition
investigated by the \ac algorithms.
Now, the set of partitions of the $n$-set $S$ is distributed in a not necessarily uniform way, but each element of $S$ has a probability of $p_i$ to belong to the partition class $C_i$ (and the probabilities $p_i$ sum up to $1$). 
%This model seems of greater significance for to real-world applications than the model from section~\ref{sec:uni}, however, it is harder to analyze
%
Our main focus here will be on the most applicable and accessible variant of the above-described model, where the number of partition class is a fixed number $k$, and $k$ is small compared to $n$. Without loss of generality, we can assume that the probabilities $p_1, p_2, \dots , p_k$ in this model satisfy $p_1\ge p_2\ge  \dots \ge  p_k$.

Intuitively, an algorithm with low average complexity should avoid making many comparisons between different classes (one such comparison is always necessary between any pair of classes, but more comparisons are not helpful). For this reason, it seems plausible that an \ac algorithm with optimal complexity should compare a new element first with the largest class identified up to this moment, as both the new element and the largest class are most likely to coincide with $C_1$, the `most likely class'. 
This is precisely how the clique \ac algorithm
from Definition~\ref{def:clique_universal} operates, where we add the additional refinement that each
new item is compared to the blocks discovered so far
in decreasing order of their sizes.
With this refinement, we conjecture the following.

%In this spirit, let us define an \ac algorithm which we will call the \emph{clique algorithm} (see Definition~\ref{def:clique_universal}), as follows. We will assume the elements of $S$ are ordered as $s_1, s_2, \dots , s_n$. The algorithm starts by comparing $s_1$ and $s_2$. Then, at each step $i<n$, it takes element $s_{i+1}$ and compares it with earlier elements until it can be identified with one of them. This is done by first comparing $s_{i+1}$ to an arbitrary element from  the largest identified class of earlier elements. If the answer is positive, $s_{i+1}$ is added to this class, otherwise we compare $s_{i+1}$ to an element from the second largest class, and so on, until we can identify $s_{i+1}$ with one class. In case of ties between sizes of classes, we arbitrarily choose one class. After adding $s_{i+1}$ to a class, we continue with element $s_{i+2}$. 
%(Note that strictly speaking, there is not just one clique algorithm, as the above procedure depends on the order of $S$, and there could be different rules for picking the element from a class that $s_{i+1}$ is compared to. However, as these variants have the same complexity, for simplicity we will just  write \emph{the} clique algorithm for any of these.)

%We conjecture the following.

\begin{conjecture}\label{clique-algo}
For an $n$-set with a random partition with probabilities $p_1, p_2, \dots , p_k$,  
the clique algorithm has  minimal average complexity among all \ac algorithms. 
\end{conjecture}
%We  believe that certain close variants of the clique algorithm  also attain the lowest average complexity. (delete?)\marginpar{ $\leftarrow$}

In support of Conjecture~\ref{clique-algo}, we present the following results. 
Firstly, we exhibit the limit behaviour of the expected number of queries of the clique algorithm. The simple proof of this theorem can be found in the appendix.
It allows one to decide for practical cases
whether direct classification by human experts
or pairwise queries will be more efficient
to annotate the training data.

\begin{theorem}\label{clique_algo_complexity}
 Let $p_1\ge\dots\ge p_k$ be fixed with $\sum_{i=1}^kp_i=1$.  Let $X$ denote the expected number of queries made by the clique algorithm with these parameters for an $n$-set.  Then
 $\mathbb E[X]\sim \sum_{i=1}^k ip_in.$
%$$\mathbb E[Y]\sim n-k+\sum_{i=2}^k (i-1)p_in.$$
%$$\mathbb E[X]\sim n-k+\sum_{i<j}\min\{p_i,p_j\}n.$$
\end{theorem}

%We can also show a lower bound on the expected complexity of any \ac algorithm in our setting.\begin{theorem}\label{lower_bound} Lower bound on the expected complexity of any \ac algorithm. \end{theorem}
%Comparing the bounds in Theorems~\ref{clique_algo_complexity} and~\ref{lower_bound}, we see that the clique algorithm is not too far from optimal.  

We now compare the complexity of the clique algorithm with the complexity of an algorithm that  chooses its queries randomly. More precisely, we define the \emph{random algorithm} to be the one that at each step $t$ compares a pair of elements chosen uniformly at random among all pairs whose relation is not known at this time (that is, they are neither known to belong to the same class, nor known to belong to different classes). The reason for analyzing the random algorithm is that one may think of this algorithm as the one that models the situation where no strategy at all is used by the human experts, which might make this procedure cheap to implement.
It turns out, however, that there is a cost in form of larger average complexity associated to the random algorithm, if compared to our proposed candidate for the optimal algorithm, the clique algorithm. 

%The limit of the average expected complexity of the random algorithm is given by  our next result.

\begin{theorem}\label{random_algo_complexity}
 Let $p_1\ge\dots\ge p_k$ be fixed with $\sum_{i=1}^kp_i=1$.   Let $X$ be the number of queries of a random algorithm with these parameters for an $n$-set.  Then
$\mathbb E[X]\sim n-k+\sum_{i<j}f(p_i,p_j)n$,
where
$$
f(\alpha,\beta)=
\begin{cases}
2\alpha &\alpha=\beta\\
\frac{2\alpha\beta\ln(\alpha/\beta)}{\alpha-\beta}&\alpha\ne\beta.
\end{cases}
$$
\end{theorem}

Note that if all the $p_i$ are equal to $1/k$, then by Theorems~\ref{clique_algo_complexity} and~\ref{random_algo_complexity},  the expected number of queries produced by the random algorithm is asymptotically $n-k+(2n/k)\binom k2=k(n-1)$, while for the clique algorithm this number is $(k+1)n/2$.

%%%%%%%%%%%%%%%%%%%%%%%%%%%%%%%%%%%%%%%%%%%%%%%%%%%%%%%%%%%
    \subsection{Noisy queries} \label{sec:noisy}
%%%%%%%%%%%%%%%%%%%%%%%%%%%%%%%%%%%%%%%%%%%%%%%%%%%%%%%%%%%

We now discuss the situation when answers to the query can be inconsistent,
due to errors from the human experts
or ambiguity in the data.
Different perspectives can be found in \cite{davidson2014top, mazumdar2017clustering}.
%So far, any two items linked by a chain of positive queries
%were indistinguishable using further queries.
%
Let $G$ denote the graph where vertices correspond to items from the data set, edges to queries, so each edge is either \emph{positive} or \emph{negative} depending on the answer.
Because of potential errors,
the aggregated graph is not sufficient for the analysis anymore,
but it can be recovered by contracting to a vertex
each positive component of $G$.

\paragraph{Correcting errors.}
An inconsistency is detected if and only if
$G$ contains a \emph{contradictory cycle}:
a cycle where all edges except one are positive.
At each step, we consider the shortest contradictory cycle
and ask a query that cuts it in two,
following a divide-and-conquer strategy.
After a logarithmic number of queries,
if no additional error occurred,
a false answer or ambiguous data is identified
and the answer is corrected.
At any point, if the number of contradictions detected
grows out of proportion (edit war between human experts),
a classic clustering algorithm can be applied
as a last resort to settle the differences of opinions.
We now focus on the problem of detecting inconsistencies.

\paragraph{Bounded number of errors.}
To ensure the detection of at most $k$ errors,
each positive component of graph $G$ must be $(k+1)$-edge-connected
and any two positive components must be linked by at least $k+1$
negative edges.
The minimal number of queries for $n$ items and $b$ blocks,
assuming no block has size $1$,
is then $(k+1) \big(\binom{b}{2} + n/2\big)$,
because each vertex of $G$ has at least $k+1$ positive neighbors.

\paragraph{Small probability of error.}
To avoid this high cost, the error detection criteria can be relaxed.
We now consider that each answer has a small probability $p$ of error
and add a few queries at the end of the \ac algorithm,
while keeping the probability of undetected errors low.
More precisely, let $c_0 + c_1 p + c_2 p^2 + \cdots$
denote the Taylor expansion of this probability at $p=0$.
Since $p$ is assumed to be small,
our aim is to minimize the vector $(c_0, c_1, \ldots)$ for the lexicographic order.
During the \ac algorithm, when a query between two classes is needed,
we choose the items in those classes to maintain the following structure.
Any positive component should be a tree
with vertices of degree at most $3$.
We call \emph{$2$-path}
a path of vertices of degree $2$ linking two vertices of degree $3$.
The $2$-paths should all have length close to a parameter $r$ of the algorithm.
At the end of the algorithm, we introduce additional queries
between pairs of leaves of the same tree,
turning each positive component of size at least $2$ into either
a $3$-edge-connected graph or a cycle of length close to $r$.
Queries are also added to ensure that any two positive components
are linked by at least $3$ negative edges.
Assuming the partition contains $b$ blocks
and the average length of the $2$-paths is $r$,
the number of additional queries compared to the noiseless setting is
approximately $\frac{n}{3r+2} + b$ inside the blocks,
plus the potential queries between blocks (at most $2 \binom{b}{2}$).
This setting ensures $c_0 = c_1 = 0$
and minimizes $c_2$ (see \cite{bauer1985combinatorial}).
If the length of all $2$-paths and
the sizes of positive components that are cycles
are bounded by $r'$,
then $c_2 = \binom{r'+1}{2} \frac{3n}{3r+2}$.
Thus, the choice of the parameter $r$ is a trade-off
between the number of additional queries and
the robustness to noise of the algorithm.

%%%%%%%%%%%%%%%%%%%%%%%%%%%%%%%%%%%%%%%%%%%%%%%%%%%%%%%%%%%
\section{Proofs} \label{sec:proofs}
%%%%%%%%%%%%%%%%%%%%%%%%%%%%%%%%%%%%%%%%%%%%%%%%%%%%%%%%%%%
%\subsection{Proofs of the results from section~\ref{sec:uni}}
%%%%%%%%%%%%%%%%%%%%%%%%%%%%%%%%%%%%%%%%%%%%%%%%%%%%%%%%%%%
\subsection{Proof of Theorem~\ref{th:optimal_chordal}}\label{sec:proof_chordal}
We will prove Theorem~\ref{th:optimal_chordal} by analyzing the types of queries made by an \ac algorithm.
%Given a  partition $P$ of an $n$-set and an \ac algorithm, we can associate a graph $G_t$ to each step $t$ of the algorithm. Namely, $G_t$ is obtained by adding an edge for each negatively answered query, and successively identifying pairs of vertices that received a positive answer. (When identifying pairs we keep all their edges.) We call $G_t$ the {\it aggregated graph} at time $t$. Note that $G_0$ is the edgeless graph on $n$ vertices, and after the last step of the algorithm, the aggregated graph is complete and of order  $|P|$. 
For this,
we classify the queries made by an \ac algorithm into three types:
\begin{itemize}
    \item {\em Core} queries are those that receive a positive answer.
    \item A query at time $t$ is {\em excessive} if it compares vertices $x$ and $y$ that are joined by an induced path in $G_t$ on an even number of vertices that alternates between two partition classes.
    \item A query at time $t$ is {\em productive} if it is neither core nor excessive.
\end{itemize}
Note that each query is of exactly one type, as excessive queries have to receive negative answers. Also note that as the core queries are the only ones that shrink $G_t$,  the number of core queries does not depend on the algorithm.

\begin{lemma}\label{kcore}
For any \ac algorithm
and any partition of a set of size $n$ containing $k$ classes,
the number of core queries is exactly $n - k$.\qed
\end{lemma}

\iffalse
\begin{proof} 
We have $|G_t|=|G_{t-1}|-1$ if the $t$th query is core; otherwise $|G_t|=|G_{t-1}|$.  Since $|G_0|=n$ and $|G_n|=k$, there must be exactly $n-k$ core queries.
\end{proof}
\fi

Next, we show that for a random partition, also the number of productive queries is, in expectation, the same for all \ac algorithms.
\begin{lemma} \label{kproductive}
Let $\mathcal P_k$ be the set of partitions of $[n]$ into exactly $k$ sets, and choose $P\in\mathcal P_k$ uniformly at random.  Then 
all algorithms have the same expected number of productive queries.
\end{lemma}

Because of space constraints, we only present a sketch of the proof of this lemma here. The  full proof of Lemma~\ref{kproductive} is in the appendix.
\begin{proof}[Sketch of the proof of Lemma~\ref{kproductive}]
Consider a partition of $[n]$ into $k$ sets $C_1, C_2, \ldots, C_k$, and let $q_{ij}$ be the number of productive queries comparing a vertex from $C_i$ with a vertex from $C_j$.
Then $\mathbb \mean[q_{ij}]=\mathbb \mean[q_{12}]$ for all $i,j$, and by linearity of expectation the expected number of productive queries is $\binom k2\mathbb \mean[q_{12}]$.  Thus, we are done if we can prove that $\mathbb \mean[q_{12}]$ is independent from  the choice of the algorithm.
It suffices to show that $\mean[q_{12} | C_1\cup C_2= S]$ is independent from the choice of algorithm because
\begin{align*} \nonumber
\mathbb \mean[q_{12}]
    & = \sum_{S \subset [n], |S|\geq2} \mean[q_{12} | C_1\cup C_2= S] \proba[C_1\cup C_2=S].
\end{align*}

In other words, we wish to understand 
the expected number of productive queries of an \ac algorithm
working on a partition with exactly two classes of a set of size $|S|$,  chosen uniformly at random (there are $2^{|S|}-2$ such partitions).
Observe that this is very similar to calculating 
the expected number of productive queries for a  uniformly-chosen partition with {\it at most} two classes (there are $2^{|S|}$ such partitions). In fact, it it not hard to see that these numbers only differ by a factor of $\frac{2^{|S|}}{2^{|S|-2}}$. So we can restrict our attention to the expectation of the latter number, which is easier to calculate, as the model is equivalent to assigning independently to each element of $S$ a value in $\{0,1\}$ uniformly at random.

More precisely, we will now argue that
the expected number of productive queries in this scenario is 
$\frac{n-1}{2}$, for each partition algorithm.
For this, note that each component of any aggregated graph 
 $G_t$ is either a single vertex or a nonempty bipartite graph. Moreover, each component has two possible colorings, and it is not hard to show by induction that these colorings are equally likely.
So, whenever the algorithm makes a query for two vertices from distinct components, answers `yes' and `no' are  equally likely. As the algorithm makes $n-1$ such queries (since each such query reduces the number of components by 1), it follows from linearity of expectation, that the expected number of productive queries is $(n-1)/2$.
\end{proof}

Finally, we analyze the excessive queries of an \ac algorithm.
\begin{lemma} \label{th:p_forest}
An \ac algorithm  makes no excessive queries 
if and only if it is chordal.
\end{lemma} 

For the proof, we need the following easy lemma whose proof can be found in %section~\ref{sec:equiv} of
the appendix. 
%%   This lemma is also used in that same section for showing the equivalence of the conditions in the def of the chordal algorithm.
\begin{lemma} \label{factChordal}
For each non-chordal \ac algorithm, there is an input partition and a time $t$ such that $G_t$ has an induced $C_4$ one of whose edges comes from a negative query in step $t-1$.
\end{lemma}

\begin{proof}[Proof of Lemma~\ref{th:p_forest}]
By definition, a chordal \ac algorithm has no aggregated graphs with induced cycles of length at least $4$. So, since an excessive query, if answered negatively, creates an induced cycle of even length at least 4, chordal algorithms make no excessive queries.

For the other direction, let us show that any non-chordal \ac algorithm makes at least one excessive query for some partition. By Fact~\ref{factChordal},  a non-chordal \ac algorithm has, for some partition of the ground set, an aggregated graph $G_t$ with an induced $C_4$, on vertices $v_1, v_2, v_3, v_4$, in this order, such that the last query concerned the pair  $v_1, v_4$.  There is a realization of $G_t$ where $v_1$ and $v_3$ are in one partition class, and $v_2$ and $v_4$ are in another partition class. For this partition, the query  $v_1, v_4$ at time $t-1$ is an excessive query.
\end{proof}

We are ready for the proof of Theorem~\ref{th:optimal_chordal}.

\begin{proof}[Proof of Theorem~\ref{th:optimal_chordal}]
By Lemmas~\ref{kcore} and~\ref{kproductive}, all \ac algorithms have  the same expected number of core queries and productive queries. So the optimal algorithms are the ones with the minimum expected number of excessive queries. By Lemma~\ref{th:p_forest}, these are the chordal algorithms.
\end{proof}

%%%%%%%%%%%%%%%%%%%%%%%%%%%%%%%%%%%%%%%%%%%%%%%%%%%%%%%%%%%

%%%%%%%%%%%%%%%%%%%%%%%%%%%%%%%%%%%%%%%%%%%%%%%%%%%%%%%%%%%

\subsection{Proof of Theorem~\ref{th:chordal_same_distribution}}\label{sec:sketch_distri}

%%%%%%%%%%%%%%%%%%%%%%%%%%%%%%%%%%%%%%%%%%%%%%%%%%%%%%%%%%%

 %For space constraints, we only give a brief sketch
 %here; the full proof of Theorem~\ref{th:chordal_same_distribution} can be %found in the appendix.
The full proof is in the appendix. We prove  a more general result which allows for the algorithm to start with any aggregated graph instead of the empty graph. 
We prove this using induction on the number of non-edges of~$G=(V,E)$, the base case being trivial. We fix some useful notation now: For $x,y\in V$, set $G(xy)=(V, E\cup \{xy\})$, and let $G_{xy}$ be  obtained from $G$ by identifying $x$ and $y$.

For the induction step  consider a graph $G$ with $k+1$ missing edges, and let $A_0$, $A_1$ be two distinct chordal \ac algorithms for $G$. 
%If their first queries are the same, say they query the edge $e$ then by induction we know that for both $G_e$ and $G(e)$, the two algorithms have the same distribution if we let them start there. As the distribution for an algorithm starting at $G$ is uniquely obtained from the complexity distributions of the same algorithm starting at  $G_e$ and at $G(e)$, we see that $A_0$ and $A_1$ have the same complexity  distribution.
By induction, we can assume that $A_0$ and $A_1$ differ in their first queries. Say the first query of $A_i$ is $u_i$, $v_i$, for $i=0,1$. Then $G(u_iv_i)$ is chordal for $i=0,1$. Note that we can assume that $u_0\neq u_1$. We distinguish two cases.

{\bf Case 1.} $G(u_0v_0)(u_1v_1)$ is chordal and
 moreover, if $v_0= v_1$ then $u_0u_1\notin E(G)$. Then, for $i=0,1$, the edge $u_iv_i$ can be chosen as the first edge of a chordal \ac algorithm for $G(u_{1-i}v_{1-i})$ or for $G_{u_{1-i}v_{1-i}}$. As the induction hypothesis applies to $G(u_{1-i}v_{1-i})$ and to $G_{u_{1-i}v_{1-i}}$, we can assume that $u_iv_i$ is the second edge in $A_{1-i}$. 
Observe that for each $i=0,1$ after the second query of $A_i$, we arrive at one of the four graphs $(G_{u_0v_0})_{u_1v_1}$, $G(u_0v_0)_{u_1v_1}$, $G(u_1v_1)_{u_0v_0}$, $G(u_0v_0)(u_1v_1)$. Thus the complexity distribution of $A_0$ and $A_1$ is identical (as it can be computed from the complexity distribution for the algorithms starting at these four graphs).

{\bf Case 2.} $G(u_0v_0)(u_1v_1)$ is chordal, $v_0=v_1$ and $u_0u_1\in E(G)$, or $G(u_0v_0)(u_1v_1)$ is not chordal. This case is harder to analyze, but one can show that either  there is an edge 
 $uv\in E(G)$ such that $G(uv)$, $G(uv)(u_0v_0)$ and $G(uv)(u_1v_1)$ are  chordal, or
 $G$ has a very specific shape. In the former case, we proceed as in the previous paragraph using the edge $uv$ as a proxy. In the latter case our argument relies on our knowledge on the structure of $G$.

%%%%%%%%%%%%%%%%%%%%%%%%%%%%%%%%%%%%%%%%%%%%%%%%%%%%%%%%%%%
\subsection{Proof of Theorem~\ref{th:chordal_limit_law}}
%%%%%%%%%%%%%%%%%%%%%%%%%%%%%%%%%%%%%%%%%%%%%%%%%%%%%%%%%%%

According to Theorem~\ref{th:chordal_same_distribution},
all chordal algorithms share the same complexity distribution,
so we investigate a specific chordal algorithm,
the \emph{universal \ac algorithm} (see Definition~\ref{def:clique_universal})
without loss of generality.
%The partition $R$, output of the algorithm, is updated progressively.
%We start with $R$ equal to the empty partition, and all items unmarked.
%At each step, consider the largest unmarked item $u$,
%compare it to all other unmarked items.
%Let $B$ denote the set of items $v$ such that the query $(u,v)$
%received a positive answer.
%Then $B \cup \{u\}$ is added to $R$ as a new block.
%The corresponding items are removed from the set of unmarked items
%and the algorithm iterates on the next largest unmarked item,
%until there are none left.
This algorithm first computes the block $B$ containing the largest item
by comparing it to all other items,
then calculates the partition $Q$ for the remaining items,
and finally inserts the block $B$ in $Q$.
Let $\query(p)$ denote the number of queries used by the universal \ac algorithm
to recover partition $p$.
Let us introduce the generating function
\[
    P(z,q) =
    \sum_{\text{partition $p$}}
    q^{\query(p)}
    \frac{z^{|p|}}{|p|!}.
\]
The \emph{symbolic method} presented in~\cite{FS09}
translates the description of the algorithm
into a the differential equation characterizing $P(z,q)$
\[
    \partial_z P(z,q) = P(q z, q) e^{q z}
\]
with initial condition $P(0, q) = 1$.
Observing that the function $e^{\frac{q}{1-q} z}$
is solution of a similar differential equation,
%\[
%    \partial_z f(z,q) = \frac{q}{1-q} f(q z, q) e^{q z},
%\]
we consider solutions of the form $P(z,q) = A(z,q) e^{\frac{q}{1-q} z}$.
The differential equation on $P(z,q)$ translates into
a differential equation on $A(z,q)$
\[
    \partial_z A(z,q) + \frac{q}{1-q} A(z,q) = A(q z, q)
\]
with initial condition $A(0,q) = 1$. Decomposing $A(z,q)$ as a series in $z$,
we find the solution
\[
    A(z,q) =
    \sum_{k \geq 0}
    \left(\frac{1-q}{q}; q \right)_k
    \frac{\left(- \frac{q}{1-q} z\right)^k}{k!}.
\]
By definition, the probability generating function $\pgf_n(q)$
of the complexity distribution when the set of items has size $n$
is linked to our generating function by the relation
\[
    \pgf_n(q) = \frac{n!}{B_n} [z^n] P(z,q),
\]
where the Bell number $B_n$ counts the number of partitions of size $n$.
The first exact expression from the theorem
%\[
%    \pgf_n(q) =
%    \frac{1}{B_n}
%    \left(\frac{q}{1-q}\right)^n
%    \sum_{k=0}^n
%    \binom{n}{k} (-1)^k
%    \left( \frac{1-q}{q}; q\right)_k
%\]
is obtained directly by coefficient extraction.
The second one requires using the following classic $q$-identities
\[
    [n]_q! = \frac{(q;q)_n}{(1-q)^n},
    \quad
    \frac{1}{(x;q)_{\infty}} =
    \sum_{n \geq 0}
    \frac{x^n}{(q;q)_n},
    \quad
    e_q(x) = ((1-q)x; q)_{\infty}^{-1}.
\]
To obtain the Gaussian limit law,
we prove that the Laplace transform of the normalized random variable $X_n^\star = (X_n - E_n) / \sigma_n$
\[
    \mean(e^{s X_n^\star}) = \pgf_n(e^{s / \sigma_n}) e^{- s E_n / \sigma_n}
\]
converges to the Laplace transform of the standard Gaussian $e^{s^2/2}$
pointwise for $s$ in a neighborhood of $0$.
To do so, we apply to the second expression the Laplace method for sums
\cite[p. 761]{FS09},
using in the process a $q$-analog of Stirling's approximation~\cite{moak1984q}.

%%%%%%%%%%%%%%%%%%%%%%%%%%%%%%%%%%%%%%%%%%%%%%%%%%%%%%%%%%%
%%%%%%%%%%%%%%%%%%%%%%%%%%%%%%%%%%%%%%%%%%%%%%%%%%%%%%%%%%%
%\subsection{Proofs of the results from section~\ref{sec:random}}
%%%%%%%%%%%%%%%%%%%%%%%%%%%%%%%%%%%%%%%%%%%%%%%%%%%%%%%%%%%

%%%%%%%%%%%%%%%%%%%%%%%%%%%%%%%%%%%%%%%%%%%%%%%%%%%%%%%%%%%
\subsection{Proof of Theorem~\ref{random_algo_complexity}}

Since Lemma~\ref{kcore} is still valid in this setting, it suffices to prove the following, with $f$ as in Theorem~\ref{random_algo_complexity}.
\begin{theorem}\label{random_algo_complexity_crossedges}
 Let $p_1\ge\dots\ge p_k$ be fixed with $\sum_{i=1}^kp_i=1$.   Let $X$ denote the number of edges between classes using a random algorithm.  Then
$\mathbb E[X]\sim \sum_{i<j}f(p_i,p_j)n.$
\end{theorem}
\begin{proof}
 We only give a sketch here, see the appendix for further detail. Instead of analyzing the random algorithm, we analyze   the following process.  Begin with all vertices marked {\em active}.  At each time step, pick (with replacement) a random pair $\{u,v\}$ and:
\begin{itemize}
    \item If $u,v$ are active and from distinct classes $i,j$ then say we have generated an {\em $ij$-crossedge}.
    \item If $u,v,$ are both active and in class $i$ then mark exactly one of $u$ and $v$ inactive.
    \item If one of $u,v$ is inactive, then do nothing.
\end{itemize}
Note that as the process runs, the number of active vertices is monotonic decreasing, and we are increasingly likely to choose pairs where one vertex is inactive.  These contribute to the new process, but do not generate new comparisons between pairs. So we are looking at a (randomly) slowed down version of the random algorithm; but this makes the analysis much simpler.

Let $x_i(t)$ denote the number of active class $i$ vertices after $t$ time steps, and $x_{ij}(t)$ denote the number of $ij$-crossedges that are generated in the first $t$ time steps.  Then at step $t+1$, the probability that we pick two active vertices in class $i$ is
$$\binom{x_i(t)}{2}\Big/\binom n2 \sim \frac{x_i(t)^2}{n^2}.$$
Writing $p=p_i$, we estimate $x_i(t)$ via a function $x=x(t)$ satisfying the differential equation
$\partial_t{x}(t) = -\frac{x(t)^2}{n^2}$, with $x(0) = pn$
This has solution
$$x(t)=\frac{n^2}{t+n/p}=\frac{pn}{1+pt/n}.$$
%Note that at time $\lambda n$, as $\lambda$ gets large, we get $$x(\lambda n)\sim\frac{pn}{1+\lambda p}.$$ 
One can show that with high probability, $x_i(t)$ tracks $x(t)$ quite closely.  

Now we estimate the number of $ij$-crossedges.  Using our estimates for $x_i(t)$ and $x_j(t)$, we see that the probability of an $ij$ crossedge at step $t+1$ is 
$$
x_i(t)x_j(t)/\binom n2
\sim \frac{2x_i(t)x_j(t)}{n^2}
\approx \frac{2n^2}{(t+n/p_i)(t+n/p_j)}.
$$
Let $p=p_i$ and $q=p_j$.   Similarly as before, we can model the growth of $x_{ij}(t)$ by a function $c=c(t)$ satisfying the differential equation
$$\partial_t{c(t)}=\frac{2n^2}{(t+n/p)(t+n/q)}$$ with $c(0)= 0$. Calculations show that if $p=q$, then
\begin{align*}
    c(t)=2pn\left(1-\frac{1}{2+2pt/n}\right)\sim 2pn
\end{align*}
and if $p\ne q$, then
\begin{align*}
    c(t)\sim \frac{2npq\ln(p/q)}{p-q}
\end{align*}
as $t/n\to\infty$.  In order to prove the theorem, we now run the process for time $Kn$, where $K$ is a large constant.  We note that at this point, we have (with high probability) at most $cn$ remaining active vertices for some small constant $c$.  But now we revert to the original process: noting that for any partition into $k$ classes, a fraction of at least (about) $1/k$ of the pairs lie inside some class, we see that at each step the process reduces the number of vertices with constant probability. The remaining expected running time is therefore $O(kcn)$.  
\end{proof}

We note that it is also possible to prove a central limit theorem for the running time of the clique algorithm (roughly: after $n^{1/2}$ comparisons, we are very likely to have seen representatives from every set in the partition, and more from the $i$th class than the $j$th class whenever $p_i>p_j$; we estimate the contribution from the remaining steps by an sum of independent random variables).  
We do not pursue the details here.

%%%%%%%%%%%%%%%%%%%%%%%%%%%%%%%%%%%%%%%%%%%%%%%%%%%%%%%%%%%
        \section{Conclusion} \label{sec:conclusion}
%%%%%%%%%%%%%%%%%%%%%%%%%%%%%%%%%%%%%%%%%%%%%%%%%%%%%%%%%%%

In this article, motivated by the building of training data for supervised learning,
we studied active clustering using pairwise comparisons
and without any other additional information (such as a similarity measure between items).

Two random models for the secret set partition where considered:
the uniform model and the bounded number of blocks model.
They correspond to different practical annotation situations.
The uniform model reflects the case where nothing is known about the set partition.
In that case, many clusters will typically be discarded at the end of the clustering process,
as they are too small for the supervised learning algorithm to learn anything from them.
Thus, this is a worst case scenario.
When some information is available on the set partition,
such as its number of blocks or their relative sizes,
the bounded number of blocks model becomes applicable.

\paragraph{Comparison between direct labeling and pairwise comparisons.}
As a practical application of this work,
we provided tools to decide whether direct labeling or pairwise comparisons
will be more efficient for a given annotation task.
One should first decide whether the uniform model
or the bounded number of blocks model best represents the data.
In both cases, our theorems provide estimates
for the number of pairwise queries required.
Then the time taken by an expert to answer
a direct labeling query or a pairwise comparison
should be measured.
Finally, combining those estimates, the time required to annotate the data
using direct labeling or pairwise comparison can be compared.
We also provided tools to detect and correct errors in the experts' answers.

\paragraph{Similarity measures.}
Generally, a similarity measure on the data is available
and improves the quality of the queries we can propose to the human experts.
This similarity measure can be a heuristic that depends only on the format of the data.
For example, if we are classifying technical documents,
a text distance can be used.
The similarity could also be trained on a small set of data already labeled.
This setting has been analyzed by \cite{mazumdar2017sideinfo}.
Our motivation for annotating data is to train a supervised learning algorithm.
However, one could use active learning to merge and replace the annotation and learning steps.
The problem is then to find the queries that will improve the learned classifier the most.

In this article, we focused on the case
where such similarity measures are not available.
However, we are confident that the mathematical tools developed here
will be useful to analyze more elaborate settings as well.
In particular, the aggregated graph contains exactly
the information on the transitive closure of the answers to the pairwise queries,
so its structure should prove relevant whenever pairwise comparisons are considered.

\paragraph{Open problems.}
We leave as open problems the proof that the clique algorithm
reaches the minimal average complexity in the bounded number of blocks model,
and the complexity analysis of the random algorithm in the uniform model.

%%%%%%%%%%%%%%%%%%%%%%%%%%%%%
~\newline

\paragraph{Acknowledgments and Disclosure of Funding.}
We thank Maria Laura Maag (Nokia) for introducing us
to the problem of clustering using pairwise comparisons,
and Louis Cohen, who could sadly not continue working with us.
The authors of this paper met through and were supported by the RandNET project
(Rise project H2020-EU.1.3.3).
Alex Scott was supported by EPSRC grant EP/V007327/1.
Quentin Lutz and \'{E}lie de Panafieu produced part of this work
at Lincs (\texttt{www.lincs.fr}). Quentin Lutz is supported by Nokia Bell Labs (CIFRE convention 2018/1648).
Maya Stein was supported by ANID Regular Grant 1180830, and by ANID PIA ACE210010.
%%%%%%%%%%%%%%%%%%%%%%%%%%%%%

%\bibliography{neurips_biblio}
\printbibliography

%%%%%%%%%%%%%%%%%%%%%%%%%%%%%%%%%%%%%%%%%%%%%%%%%%%%%%%%%%%%

\appendix

\section{Appendix}

\subsection{Equivalence in the definition of chordal algorithms}\label{sec:equiv}

%%%%%%%%

In this short section we show the equivalence of the conditions in the definition of chordal \ac algorithms from section~\ref{sec:uni}, and some related easy lemmas. 
We start by proving a lemma from section~\ref{sec:proof_chordal}. 
%\begin{lemma}\ref{factChordal}
%For each non-chordal \ac algorithm, there is an input partition and a time $t$ such that $G_t$ has an induced $C_4$ one of whose edges comes from a negative query in step $t-1$.
%\end{lemma}
\begin{proof}[Proof of Lemma~\ref{factChordal}]
Since the algorithm is not chordal, for some  input partition one of the aggregated graphs $G_t$ of the \ac algorithm has an induced cycle $C$ of length at least $4$. Consider a realization of $G_t$ where the vertices of $C$ are all in distinct partition classes. Then there is a time $t'$ when four of the vertices of $C$ form an induced $C_4$ in $G_{t'}$.
\end{proof}

The next lemma is used both in the proof of the equivalence of the conditions in the definition of chordal \ac algorithms, and in section~\ref{sec:distri}. Recall that for a graph $G=(V,E)$ we write $G(uv)=(V, E\cup \{uv\})$.
\begin{lemma} \label{th:chordality_pair_vertices}
Let $G$ be a chordal graph,  let $u, v\in V(G)$
be distinct and  nonadjacent, and assume $G(u,v)$ is chordal.
Then
$N(u)\cap N(v)$ 
is complete and separates $u$ and $v$ (in $G$).
\end{lemma}
\begin{proof}
Note that   $N(u)\cap N(v)$ is  complete, as otherwise there are two non-adjacent vertices $x,y\in N(u)\cap N(v)$, and $(u, x, v, y)$ is an induced cycle in $G$, a contradiction. It remains to show that $N(u)\cap N(v)$ separates $u$ from $v$. Assume this is not the case, then there is an induced path $P=(u, p_1, ..., p_k, v)$ that avoids $N(u)\cap N(v)$. In particular, $k\ge 2$. Adding the edge $uv$ to $P$ we obtain an induced cycle of length at least $4$, a contradiction to $G(u,v)$ being chordal.
\end{proof}

Now we prove the equivalence of the three conditions.

\begin{lemma} \label{lem:equiv}
The following conditions are equivalent for any \ac algorithm.
\begin{enumerate}[(i)]
\item\label{chordalAproof}
for all input partitions, each aggregated graph is chordal,
\item\label{chordalCproof} for all input partitions, no aggregated graph has an induced $C_4$,
\item\label{chordalBproof} for all input partitions, for each query  $u$, $v$, the intersection of the neighborhoods of $u$ and $v$  separates $u$ and $v$. 
\end{enumerate}
\end{lemma}
\begin{proof}
By Lemma~\ref{factChordal}, we know that~\eqref{chordalCproof} implies~\eqref{chordalAproof}, and by
 Lemma~\ref{th:chordality_pair_vertices} we know  that~\eqref{chordalAproof} implies~\eqref{chordalBproof}. So we only need to show that~\eqref{chordalBproof} implies~\eqref{chordalCproof}. For this, assume there is an input partition, and a  time $t$, such that the aggregated graph $G_t$ at time $t$ has an induced cycle $(v_1, v_2, v_3, v_4)$. We can assume that $t$ is the first such time. Then either $G_t$ arose by adding an edge of this cycle, say the edge $v_1v_4$, or by identifying two vertices $x$ and $y$ to a new vertex from the cycle, say $v_1$. In the first case, the query $v_1$, $v_4$ in step $t-1$ did not meet
the requirement in~\eqref{chordalBproof}, a contradiction.
 In the second case, $x$ and $y$ are joined by the induced path $(x, v_2, v_3, v_4, y)$ or $(y, v_2, v_3, v_4, x)$, so the query $x$, $y$ in step $t-1$ did not meet
the requirement in~\eqref{chordalBproof}, a contradiction.
\end{proof}

%%%%%%%%%%%%%%%%%%%%%%%%%%%%%
%%%%%%%%%%%%%%%%%%%%%%%%%%%%%

\subsection{Full proof of Lemma~\ref{kproductive}}

%%%%%%%%%%%%%%%%%%%%%%%%%%%%%

Let ${\mathcal P}_2(n)$ denote the family of partitions of size $n$
where each element receives a random value in $\{0,1\}$ uniformly and independently,
so that two elements belong to the same block
if and only if they share the same value.
Thus, all partitions in ${\mathcal P}_{2}(n)$ have one or two blocks.

\begin{lemma} \label{th:p_lower_bound}
The expected number of productive queries made by any \ac algorithm
on a random partition from $\mathcal P_2(n)$ is exactly 
$\frac{n-1}{2}$.
\end{lemma}

\begin{proof}
Let us consider the (random) sequence of graphs $G_1,\dots,G_n$ produced by running the 
algorithm on a random element of ${\mathcal P}_2(n)$.  As core queries result in contractions, it follows that (for every $t$) every component of $G_t$ is either a single vertex or a nonempty bipartite graph.  Note that for each component there are two possible colorings; thus the number of possible colorings of $G_i$ is $2^{\kappa(G_i)}$, where $\kappa$ denotes the number of components.

We prove by induction that: 
\begin{itemize}
    \item For each $i$, if the $i$th query joins two components of $G_{i-1}$ then it is productive with probability half.
    \item The $2^{\kappa(G_i)}$ colorings of $G_i$ are equally likely.
\end{itemize}

The second property holds for $G_0$, so it is enough to prove the inductive step.
Consider the query made at stage $t$, say joining vertices $x$ and $y$.  If $x$ and $y$ lie in the same component of $G_{t-1}$ then the query is excessive, the components do not change, and the two bullets hold.  Thus we may restrict our attention to queries such that $x$ and $y$ lie in distinct components, say $H_x$ and $H_y$.  There are four possible colorings of $H_x\cup H_y$.  The four colorings are equally likely and are independent from the coloring of the rest of the graph.  It is now easily checked that $x$ and $y$ have the same color with probability $1/2$, and so the probability that the query is productive is $1/2$.  Furthermore, adding the edge $xy$ joins $H_x$ and $H_y$ into a single component, and the two possible colorings of this component are equally likely (and remain independent from the coloring of the rest of $G$).  Thus the two bullets hold, and the induction is complete.

There are in total $n-1$ steps at which the algorithm makes a query joining distinct components (as each such query reduces the number of components by 1).  So, by linearity of expectation, the expected number of productive queries is $(n-1)/2$.
\end{proof}

\begin{lemma} \label{th:two_blocks_lower_bound}
The expected number of productive queries of an \ac algorithm
working on a partition of a set $S$
containing exactly two blocks chosen uniformly at random is
exactly 
\[
    \frac{2^{|S|}}{2^{|S|} - 2} \frac{|S|-1}{2}.
\]
\end{lemma}

\begin{proof}
Consider an \ac algorithm, and let $\alpha (|S|)$ denote the expected number of productive queries.  Now run the algorithm on a partition from ${\mathcal P}_2(|S|)$.  If the algorithm is fed one of the two constant colorings then it makes exactly $|S|-1$ queries, all of which are core (and therefore not productive).  %[Note that it is important that the algorithm can't use the fact that there are two colors to stop if it reaches a state with just two isolated vertices: it's required to produce a certificate that the partition is good, and so must compare the last two vertices.]

The probability that a partition $P\in{\mathcal P}_2(n)$ is constant is $2/2^{|S|}$; and if we condition on $P$ being nonconstant then it is uniformly distributed among the set of partitions with exactly two blocks.  By Lemma \ref{th:p_lower_bound}, we conclude that the expected number of productive queries satisfies
$$\frac{|S|-1}{2}=
\alpha(|S|)\proba[\mbox{$P$ nonconstant}] + 0\proba[\mbox{$P$ constant}]=\frac{2^{|S|}-2}{2^{|S|}}\alpha(|S|).
$$
The result follows by solving for $\alpha(|S|)$.
\end{proof}

We are now ready for the full proof of Lemma~\ref{kproductive}.
\begin{proof}[Proof of Lemma~\ref{kproductive}]
Fix $k$ and consider a partition of $[n]$ into exactly $k$ sets $C_1, C_2, \ldots, C_k$. 
Let $\alpha_{ij}$ denote the expected number of productive queries that compare a vertex from $C_i$ with a vertex from $C_j$.  Then all $\alpha_{ij}$ are equal, and by linearity of expectation the expected number of productive queries is $\binom k2\alpha_{12}$.  Thus it is enough to prove that $\alpha_{12}$ does not depend on the choice of algorithm.

 Let $q_{12}$ be the number of productive queries comparing a vertex from $C_1$ with a vertex from $C_2$. (so $\alpha_{12}=\mathbb E q_{12}$)
Then, considering the set $S=C_1\cup C_2$, and applying Lemma \ref{th:two_blocks_lower_bound}, we obtain
\begin{align} \nonumber
\alpha_{12}
    & = \sum_{S \subset [n], |S|\geq2} \mean[q_{12} | C_1\cup C_2= S] \proba[C_1\cup C_2=S]\\
    & = \sum_{S \subset [n], |S|\geq2} \frac{|S|-1}{2} \frac{2^{|S|}}{2^{|S|} - 2} \proba[C_1\cup C_2=S]. \label{eq:negative_complexity_lower_bound}
\end{align}
The last line is independent from the choice of algorithm,
which concludes the proof.
\end{proof}

%%%%%%%%%%%%%%%%%%
%%%%%%%%%%%%%%%%%%
\subsection{Proof of Theorem~\ref{th:chordal_same_distribution}}\label{sec:distri}
%%%%%%%%%%%%%%%%%%
This section is devoted to the proof of Theorem~\ref{th:chordal_same_distribution}. For this we will need several auxiliary lemmas. %Also, some notation will be helpful: For a graph $G=(V,E)$ let $G(uv)=(V, E\cup \{uv\})$, and let $G_{uv}$ be the graph obtained from $G$ by identifying $u$ and $v$.
We also recall a useful notation introduced earlier:
For a graph $G=(V,E)$ let $G(uv)=(V, E\cup \{uv\})$, and let $G_{uv}$ be the graph obtained from $G$ by identifying $u$ and $v$.
 
The first of our lemmas shows that every chordal graph can `grow' an edge while staying chordal.
\begin{lemma}\label{chordaledge}
Let $H$ be a chordal graph that is not complete. Then $H$ has a non-edge $e$ such that $H(e)$ is chordal.
\end{lemma}
\begin{proof}
Let $u$ be a non-universal vertex of $H$. Among all non-neighbors of $u$, choose $p_1$ such that $|N(u)\cap N(p_1)|$ is maximized. We claim that $H(up_1)$ is chordal.

Indeed, otherwise there is an induced cycle
$C = (u, p_1,\dots,p_k,u)$, with $k\ge 3$. As $p_k\in N(u)\cap N(p_{k-1})\setminus N(p_1)$, our choice of $p_1$ guarantees that there is a vertex $w\in N(u)\cap N(p_1)\setminus N(p_{k-1})$. 
Let $j\le k-2$ be the largest index in $[k-2]$ such that $wp_j\in E(H)$.
Then, depending on whether the edge $wp_k$ is present, either $(w,p_j,\dots,p_k,u, w)$ or $(w, p_j,\dots,p_k,w)$ is an induced cycle of length at least 4 in $H$, a contradiction.
\end{proof}

Our next two lemmas are more technical. They give a structural characterization of those aggregated  graphs where consecutive queries cannot easily be interchanged. In order to make their statement easier, let us say that a graph $G$ has a \emph{complete separation $(A,K,B)$} if
$V(G)$ is the disjoint union of $A, B, K$ so that $A\neq\emptyset\neq B$, each of $A\cup K$ and $B\cup K$ is complete, and there are  no edges from $A$ to $B$. (Observe that we allow $K$ to be empty.)

\begin{lemma}\label{compareORstructure}
Let $G$ be a chordal graph that does not have a complete separation. For $i=0,1$ let $u_iv_i$ be a non-edge of $G$ such that $G(u_iv_i)$ is chordal, and $G(u_0v_0)(u_{1}v_{1})$ is non-chordal. Then  there is a  non-edge $uv$ of $G$ such that $G(uv)$ is chordal, and $G(u_iv_i)(uv)$ is chordal for $i=0,1$. 
\end{lemma}

\begin{lemma}\label{triangle-caseII}
Let $G$ be a chordal graph that does not have a complete separation. Let $u_0, u_1, v\in V(G)$ such that for $i=0,1$, we have $u_iv\notin E(G)$ and $G(u_iv)$ is chordal, and moreover, $u_0u_1\in E(G)$. Then  there is a  non-edge $uw$ of $G$ such that $G(uw)$ is chordal, and $G(u_iv_i)(uw)$ is chordal for $i=0,1$.
\end{lemma}
The proofs of these two lemmas rely on purely structural arguments and will be postponed to the end of the section.

We are now ready to prove Theorem~\ref{th:chordal_same_distribution}. Actually, we will prove  a more general result which allows for the algorithm to start with any aggregated graph instead of starting with the empty graph. More precisely, if $G$ is an aggregated graph at time $t$ for some \ac algorithm for an $n$-set $S$, then we call the restriction of the algorithm to all queries after time $t$ that eventually lead to a realization of $G$ a \emph{\ac algorithm  starting at $G$}. 
We define the \emph{complexity distribution} of this algorithm analogously to our earlier definition. In particular, if $G$ is complete, then the complexity distribution is $(a_0, a_1, a_2, \dots, a_{\binom{n}{2}})=(1, 0, 0, \dots, 0)$.

\begin{theorem}
  For any $G$, all chordal \ac algorithms  starting at $G$ have the same  complexity distribution.
\end{theorem}
\begin{proof}
We proceed by induction on the number of non-edges of~$G$. Proving the base case is trivial as, starting from a complete graph, the only \ac algorithm has  distribution
$(a_0, a_1, a_2, \dots, a_{\binom{n}{2}})=(1, 0, 0, \dots, 0)$.

For the induction step assume that for any chordal graph with $k$ or less missing edges, all chordal \ac algorithms have the same complexity distribution, and consider a graph $G$ with $k+1$ missing edges.  Let $A_0$, $A_1$ be two distinct chordal \ac algorithms for $G$. If their first queries are the same, say they query the edge $e$, then by induction we know that for both $G_e$ and $G(e)$, the two algorithms have the same distribution if we let them start there. As the distribution for an algorithm starting at $G$ is uniquely obtained from the complexity distributions of the same algorithm starting at  $G_e$ and at $G(e)$, we see that $A_0$ and $A_1$ have the same complexity  distribution.

So we can assume that $A_0$ and $A_1$ differ in their first queries. Say the first query of $A_i$ is $u_i$, $v_i$, for $i=0,1$. Then $G(u_iv_i)$ is chordal for $i=0,1$. Note that we can assume that $u_0\neq u_1$. We will distinguish two cases.

First, let us assume that $G(u_0v_0)(u_1v_1)$ is chordal and
 moreover, if $v_0= v_1$ then $u_0u_1\notin E(G)$. Then, for $i=0,1$, the edge $u_iv_i$ can be chosen as the first edge of a chordal \ac algorithm for $G(u_{1-i}v_{1-i})$ or for $G_{u_{1-i}v_{1-i}}$. As the induction hypothesis applies to $G(u_{1-i}v_{1-i})$ and to $G_{u_{1-i}v_{1-i}}$, we can assume that $u_iv_i$ is the second edge in $A_{1-i}$. 
Observe that for each $i=0,1$ after the second query of $A_i$, we arrive at one of the four graphs $(G_{u_0v_0})_{u_1v_1}$, $G(u_0v_0)_{u_1v_1}$, $G(u_1v_1)_{u_0v_0}$, $G(u_0v_0)(u_1v_1)$. Thus the complexity distribution of $A_0$ and $A_1$ is identical (as is can be computed from the complexity distribution for the algorithms starting at these four graphs).

Now, let us assume that either $G(u_0v_0)(u_1v_1)$ is chordal, $v_0=v_1$ and $u_0u_1\in E(G)$, or $G(u_0v_0)(u_1v_1)$ is not chordal. Then, by Lemmas~\ref{compareORstructure} and~\ref{triangle-caseII}, we know that either there is an edge 
 $uv\in E(G)$ such that $G(uv)$, $G(uv)(u_0v_0)$ and $G(uv)(u_1v_1)$ are  chordal, or
$V(G)$ can be partitioned into three sets, $A$, $B$ and $K$, such that $A\cup K$ and $B\cup K$ are  complete and $K$ separates $A$ from $B$. If the former is the case, we can proceed as in the previous paragraph 
to see that every chordal \ac algorithm starting with $u_0v_0$ has the same complexity distribution as any of the chordal \ac algorithms starting with $uv$ (note that such algorithms exist by Lemma~\ref{chordaledge}), which, in turn, has the same complexity distribution as any of the chordal \ac algorithms starting with $u_1v_1$, leading to the desired conclusion. 

So assume there are sets $A$, $B$ and $K$ as above. By symmetry, we can assume that $u_0, u_1\in A$ and $v_0, v_1\in B$.
Consider the automorphism $\sigma$ of $G$ that maps $u_0$ to $u_1$ and $v_0$ to $v_1$ while keeping all other vertices fixed. We can now view $A_0$ as an algorithm  in $\sigma(G)$ that starts with the edge $u_1v_1$. By the induction hypothesis, we conclude that $A_1$ (for  $G$) has the same distribution as $A_0$ (for $\sigma(G)$, and thus also for~$G$).
\end{proof}

%%%
It remains to prove Lemmas~\ref{compareORstructure} and~\ref{triangle-caseII}.
We start by giving a characterization of graphs that remain chordal when we add either one of two edges, but not if we add both.

\begin{lemma} \label{th:four_nonchordal_vertices}
Let $G$ be a chordal graph and let $u_0, u_1, v_0, v_1\in V(G)$ 
such that  for $i=0,1$, vertices  $u_0, u_1, v_i$ are all distinct,  $u_iv_i\notin E(G)$, and
  $G(u_i v_i)$ is chordal. If $G(u_0 v_0)(u_1 v_1)$ is not chordal,
then $K:=N(u_0)\cap N(u_1)\cap N(v_0)\cap N(v_1)$ is complete, and $G-K$ has two distinct components $A$ and $B$, such that either $u_0, u_1\in A$ and $v_0, v_1\in B$, or $u_0, v_1\in A$ and $u_1, v_0\in B$.
\end{lemma}
\begin{proof}
As $G(u_0 v_0)(u_1 v_1)$ is not chordal, we know that $G(u_0 v_0)(u_1 v_1)$ has an induced cycle $C=(u_1, v_1, ..., v_{\ell-1}, v_{\ell}, ..., v_k, u_1)$, with $k\ge \ell\ge 2$, where  either $v_{\ell -1}=v_0$ and $v_\ell=u_0$, or $v_{\ell -1}=u_0$ and $v_\ell=v_0$.
According to Lemma~\ref{th:chordality_pair_vertices}, since $G(u_i v_i)$ is chordal, the set $K_i:=N(u_i)\cap N(v_i)$
is complete and separates $u_i$ and $v_i$ in $G$, 
for each $i \in \{0,1\}$.
Since $C$ has at least four vertices, and $K_i$ has neighbors $u_i, v_i$, we know that $V(C)\cap K_i=\emptyset$,  for $i=0,1$. 

 Assume there is a vertex $x\in K_0\setminus K_{1}$. Then $(v_1, ..., v_{\ell-1}, x, v_{\ell}, ..., v_k, u_1)$ is a path in $G-K_1$, in contradiction to the fact that $K_1$ separates $u_1$ from $v_1$. So $K_0\subseteq K_1$, and with the help of a symmetric argument we see that  $K_0=K_1$.
In order to finish the proof it suffices to note that the paths $(v_1, ..., v_{\ell-1})$ and $(v_{\ell}, ..., v_k)$ ensure that there are components $A$ and $B$ as desired.
\end{proof}

We now see that a graph that is obtained by gluing two graphs along a complete subgraph is chordal if and only if the two smaller graphs are.
\begin{lemma}\label{subgraph_chordality}
Let $G$ be a graph, let $A,B,K$ be a partition of  $V(G)$ such that $K$ is complete
and  there are no edges between $A$ and $B$.
 Then $G$ is chordal if and only if $G[A\cup K]$ and $G[K\cup B]$ are both chordal.
\end{lemma}
\begin{proof}
As induced subgraphs of chordal graphs are chordal, we only need to show that if both $G[A\cup K]$ and $G[K\cup B]$ are both chordal, then also $G$ is. For this, it suffices to observe that any cycle of $G$ that contains vertices from both $A$ and $B$ has to pass twice through $K$.
\end{proof}

We are now ready to prove Lemmas~\ref{compareORstructure} and~\ref{triangle-caseII}.

\begin{proof}[Proof of Lemma~\ref{compareORstructure}]
  Use Lemma~\ref{th:four_nonchordal_vertices} to see that the intersection $K$ of the neighborhoods
of $u_0$, $v_0$, $u_1$, $v_1$ is either a clique or empty, and $G-K$ has
  two connected components $A$, $B$
such that  $u_0, u_1\in A$
and $v_0, v_1\in B$ (after possibly changing the roles of $u_0$ and $v_0$).

 As $G$ has no complete separation, and as there are  no edges from $A$ to $B$,  one of $A\cup K$, $B\cup K$ has to have a non-edge; because of symmetry we can assume this is~$A\cup K$. According to Lemma~\ref{subgraph_chordality}, the subgraph of $G$ induced by $A\cup K$ is chordal. Then, according to Lemma~\ref{chordaledge}, there is also a non-edge $uv$ with $u,v\in A\cup K$ having the additional property that $G(uv)$ is chordal. As $K$ is complete, we can assume that $u\in A$.

If there is no non-edge $uv$ as desired,
we have that $G(u_iv_i)(uv)$ is non-chordal for some $i\in\{0,1\}$; by symmetry, let us assume $G(u_1v_1)(uv)$ is non-chordal.
So, we may apply Lemma~\ref{th:four_nonchordal_vertices} to see that the intersection $K'$ of the neighborhoods
of $u$, $v$, $u_1$, $v_1$ is either a clique or empty, and $G-K'$ has
  two connected components $A'$, $B'$
such that  $u, u_1\in A'$
and $v, v_1\in B'$ (after possibly changing the roles of $u$ and $v$). Note that $K'\subseteq N(u_1)\cap N(v_1)\subseteq K$. Furthermore, $K\subseteq K'$, since  $K'$ separates $u_1$ from $v_1$ and $K \subseteq N(u_1)\cap N(v_1)$. So $K=K'$. 

In particular, $v\notin K$, that is, $v\in A$. So, as $v_1\in B$, we know that $v, v_1$ lie in distinct components of $G-K$. However,  we also have that $v, v_1$ belong to the same component  (namely, $A$) of $G-K'=G-K$, a contradiction. So the desired non-edge $uv$ exists.
\end{proof}

\begin{proof}[Proof of Lemma~\ref{triangle-caseII}]
We start by proving that $N(u_0)\cap N(v)=N(u_1)\cap N(v)$. For this assume there is an $i\in\{0,1\}$ and a vertex $x\in  (N(u_{1-i})\cap N(v))\setminus N(u_i)$. Then $(u_{1-i}, x, v, u_i, u_{1-i})$ is an induced cycle of length $4$ in $G(u_iv)$, a contradiction since this graph is chordal. This proves the equality, and we set $K:=N(u_0)\cap N(v)=N(u_1)\cap N(v)$.

Because of Lemma~\ref{th:chordality_pair_vertices}, $K$ is complete and  separates $u_0, u_1$ from $v$. 
Since $G$ does not have a complete separation,  at least one of $G[A\cup K]$, $G[B\cup K]$ is not complete, but by Lemma~\ref{subgraph_chordality} both are chordal. So by Lemma~\ref{chordaledge} and, again, Lemma~\ref{subgraph_chordality}, there is a non-edge $uw$ such that $G(uw)$ is chordal. 
Now, if $uw$ is not as desired, say because $G(u_0v)(uw)$ is non-chordal, then there is an induced cycle $C$ of length at least 4  going through both $uw$ and $u_0v$. However, $C$ has to meet $K$, which implies $C$ is a triangle, a contradiction.
\end{proof}

%%%%%%%%%%%%%%%%%%%%%%%%
%%%%%%%%%%%%%%%%%%%%%%%%
%%%%%%%%%%%%%%%%%%%%%%%%
\subsection{Proof of Theorem~\ref{th:chordal_limit_law}}

%%%%%%%%%%%%
    \paragraph{Symbolic method.}
%%%%%%%%%%%%
To any sequence of numbers can be associated a generating function.
For example, consider the sequence $(B_n)_{n \geq 0}$,
where the Bell number $B_n$ denotes the number of partitions of size $n$.
The exponential generating function of partitions is then defined as
\[
    P(z) =
    \sum_{n \geq 0} B_n \frac{z^n}{n!}.
\]
The sum can be expressed at the partition level as well
\[
    P(z) =
    \sum_{p \in \partitions(n)} \frac{z^{|p|}}{|p|!}.
\]
The \emph{symbolic method}, presented in~\cite{FS09},
translates combinatorial descriptions into generating function equations.
For example, since a partition is a set of nonempty sets,
the exponential generating function of partitions is equal to
\[
    P(z) = e^{\exp(z) - 1}.
\]
The reader unfamiliar with the symbolic method can verify this result
by working on recurrences at the coefficient level.
In this example, choosing a partition of size $n$
is equivalent with choosing its number of blocks $k$,
the size $n_j \geq 1$ of the $j$th block for each $1 \leq j \leq k$,
and finally the content of those blocks, so
\[
    B_n =
    \sum_{k = 0}^n
    \sum_{\substack{n_1 + \cdots + n_k = n\\ \forall j,\ n_j \geq 1}}
    \frac{1}{k!}
    \binom{n}{n_1, \ldots, n_k}.
\]
Multiplying by $z^n / n!$, summing over $n$ and reorganizing the terms,
we indeed recover
\[
    \sum_{n \geq 0} B_n \frac{z^n}{n!} = e^{\exp(z) - 1}.
\]
In the rest of the combinatorial proofs,
the symbolic method will be preferred
and we will let the motivated reader translate those proofs
at the recurrence level.

%%%%%%%%%%%%
    \paragraph{$q$-analogs.}
%%%%%%%%%%%%
Several families of integer identities have been generalized
by introducing \emph{$q$-analogs}.
An introduction can be found in~\cite{qhistory}.
The $q$-analog of integer $n$ is defined as
\[
    [n]_q = 1 + q + \cdots + q^{n-1} = \frac{1-q^n}{1-q}.
\]
The $q$-factorial of the integer $n$ is
\[
    [n]_q! = \prod_{j=1}^n [j]_q.
\]
The $q$-exponential is defined as
\[
    e_q(z) =
    \sum_{n \geq 0}
    \frac{z^n}{[n]_q!}.
\]
The $q$-Pochhammer symbol is defined as
\[
    (a; q)_n =
    \prod_{k=0}^{n-1}
    (1 - a q^k)
\]
Observe that the $q$-analog reduce to their classic counterparts for $q=1$.

\begin{lemma} \label{th:q_identities}
The $q$-analogs satisfy the following classic identities
\[
    [n]_q! = \frac{(q;q)_n}{(1-q)^n},
    \qquad
    \frac{1}{(x;q)_{\infty}} =
    \sum_{n \geq 0} \frac{x^n}{(q;q)_n},
    \qquad
    e_q(x) =
    ((1-q) x; q)_{\infty}^{-1}.
\]
\end{lemma}

%%%%%%%%%%%%
    \paragraph{Characterizing the complexity generating function.}
%%%%%%%%%%%%

\begin{figure} 
\begin{verbatim}
def universal_ac(element_set):
    if is_empty(element_set):
        return EMPTY_PARTITION
    u = element_set.pop()
    block = {u}
    for v in element_set:
        if query(u, v):
            block.add(v)
            element_set.remove(v)
    partition = universal_ac(element_set)
    partition.add_block(block)
    return partition
\end{verbatim}
\caption{The universal \ac algorithm considers
an element, compare it to the other elements to find its block,
then partition the remaining elements.
It is named after the graph theory convention to call ``\emph{universal}''
a vertex linked to all vertices of a graph.}
\label{fig:universal_partitioning_algorithm}
\end{figure}

According to Theorem~\ref{th:chordal_same_distribution},
all chordal \ac algorithms share the same distribution
on the number of queries for random partitions of size $n$ chosen uniformly.
Thus, we study one particular chordal algorithm:
the \emph{universal \ac algorithm},
presented in Figure~\ref{fig:universal_partitioning_algorithm}.
Let $\partitions(n)$ denote the set of all partitions of size $n$,
$|p|$ the size of the partition $p$,
and $\queries(p)$ the number of queries
used by the universal \ac algorithm
to reconstruct the partition $p$.

\begin{theorem} \label{th:diff_eq}
The generating function
\[
    P(z, q) =
    \sum_{p \in \partitions(n)}
    q^{\queries(p)}
    \frac{z^{|p|}}{|p|!}
\]
is characterized by the differential equation
\[
    \partial_z P(z, q) = P(q z, q) e^{q z}
\]
and the initial condition $P(0, q) = 1$.
\end{theorem}

\begin{proof}
Consider a partition $p$ of size $n$.
Let $b$ denote the set of all elements of the block of $p$ containing $n$,
except $n$.
Let $r$ denote the partition $p$ without the block containing $n$.
Then $p$ can be recovered from the pair $(r, b)$ as follows.
The size $n$ of $p$ is $|r| + |b| + 1$,
so the block containing $n$ in $p$ was $b \cup \{n\}$
and adding this block to $r$ recovers $p$.
We have just proven that this construction is a bijection
between the partitions and the relabeled pairs
containing a partition and a set.
This bijection translates into the following identity
on the generating function $P(z)$ of partitions
\[
    \partial_z P(z) = P(z) e^z.
\]
This is no surprise, as we already know the expression of this generating function
\[
    P(z) = e^{\exp(z) - 1}
\]
from the paragraph on the symbolic method, and it indeed
satisfies this differential equation.
However, the same approach is useful to study the generating function
of the number of queries used by the universal \ac algorithm.

Consider a partition $p$ and its decomposition as a pair $(r, b)$
described above.
The universal \ac algorithm starts by comparing
one element, which is assumed to have the largest label without loss of generality,
to the other elements.
This requires $|r| + |b|$ queries.
Then the algorithm is called recursively on the partition $r$.
Thus, the generating function of partitions
with an additional parameter $q$ marking the number of queries
used by the universal \ac algorithm is
characterized by the differential equation
\[
    \partial_z P(z, q) = P(q z, q) e^{q z}.
\]
If the initial partition is empty, then there are no queries to ask,
which implies the initial condition $P(0, q) = 1$.
\end{proof}

%%%%%%%%%%%%
    \paragraph{Exact expressions.}
%%%%%%%%%%%%

The following theorem provides a solution
for the differential equation from last Theorem.

\begin{theorem}
Let  $\poch(z,a,q)$ denote the exponential generating function
associated to the $q$-Pochhammer symbol
\[
    \poch(z,a,q) =
    \sum_{k \geq 0}
    (a; q)_k \frac{z^k}{k!},
\]
then the generating function of the universal \ac algorithm complexity
is
\[
    P(z,q) =
    \poch \left( - \frac{q}{1-q} z, \frac{1-q}{q}, q\right)
    e^{\frac{q}{1-q} z}
\]
\end{theorem}

\begin{proof}
The function $f(z,q) = e^{\frac{q}{1-q} z}$
satisfies a similar differential equation
\[
    \partial_z f(z,q) = \frac{q}{1-q} f(q z,q) e^{q z}.
\]
Thus, we investigate solutions of the differential equation
from Theorem~\ref{th:diff_eq}
of the form $P(z,q) = A(z,q) e^{\frac{q}{1-q} z}$.
The differential equation on $P(z,q)$ implies
the following differential equation for $A(z, q)$
\[
    \partial_z A(z,q) + \frac{q}{1-q} A(z,q) = A(q z, q).
\]
with initial condition $A(0,q) = 1$.
Decomposing $A(z,q)$ as a series in $z$
\[
    A(z,q) =
    \sum_{k \geq 0} a_k(q) \frac{z^k}{k!},
\]
we obtain a recurrence on the $a_k(q)$
\[
    a_{k+1}(q) = - \frac{q}{1-q} a_k(q) + q^k a_k(q)
    = - \frac{q}{1-q} \left(1 - (1-q) q^{k-1}\right) a_k(q),
\]
with $a_0(q) = 1$.
We deduce
\[
    a_k(q) =
    \left( - \frac{q}{1-q} \right)^k
    \prod_{j=0}^{k-1}
    \left(1 - \frac{1-q}{q} q^j\right)
    =
    \left( - \frac{q}{1-q} \right)^k
    \left(\frac{1-q}{q}; q\right)_k.
\]
To conclude, we observe that
$\poch \left( - \frac{q}{1-q} z, \frac{1-q}{q}, q\right) e^{\frac{q}{1-q} z}$
is indeed solution of the differential equation
characterizing $P(z,q)$.
\end{proof}

Extracting the coefficient $n! [z^n]$ from the solution,
we obtain a first exact expression for $P_n(q)$ in the following Theorem,
and another one will be provided in Theorem~\ref{th:second_exact_Pnq}.
This first expression is well suited for exact computations using
a computer algebra system (we used~\cite{sagemath} to verify our calculations).
In particular, the $k$th factorial moment
of the random variable $X_n$ counting the number of queries
used by the universal \ac algorithm
on partitions of size $n$ chosen uniformly at random is
\[
    \mean(X_n (X_n - 1) \cdots (X_n - k + 1)) =
    \frac{1}{B_n} \partial_{q=1}^k P_n(q).
\]

\begin{theorem}
The generating function of the number of queries
used by the universal \ac algorithm
on partitions of size $n$ is
\[
    P_n(q) =
    \left( \frac{q}{1-q} \right)^n
    \sum_{k \geq 0}
    \binom{n}{k}
    (-1)^k
    \left( \frac{1-q}{q}; q \right)_k.
\]
\end{theorem}

We provide a second expression for the complexity generating function,
more elegant and better suited for asymptotics analysis.
It is a $q$-analog of the following classic formula
for the Bell numbers,
which counts the number of partitions of size $n$
\[
    B_n = \frac{1}{e} \sum_{m \geq 0} \frac{m^n}{m!}.
\]
This formula is obtained from the generating function
of partitions $P(z) = e^{\exp(z) - 1}$
\[
    B_n =
    n! [z^n] e^{\exp(z) - 1}
    = \frac{n!}{e} [z^n] e^{\exp(z)}
    = \frac{n!}{e} [z^n] \sum_{m \geq 0} \frac{e^{m z}}{m!}
    = \frac{1}{e} \sum_{m \geq 0} \frac{m^n}{m!}.
\]

\begin{theorem} \label{th:second_exact_Pnq}
The generating function of the number of queries
used by the universal \ac algorithm
on partitions of size $n$ is
\[
    P_n(q) =
    \frac{1}{e_q(1/q)}
    \sum_{m \geq 0}
    \frac{[m]_q^n}{[m]_q!}q^{n-m}
\]
The sum converges for $q > 1/2$.
\end{theorem}

\begin{proof}
The $q$-Pochhammer generating function is rewritten as
\begin{align*}
    \poch(z,a,q) &=
    \sum_{k \geq 0} (a; q)_k \frac{z^k}{k!}
    \\&=
    (a; q)_\infty
    \sum_{k \geq 0} \frac{1}{(a q^k; q)_\infty} \frac{z^k}{k!}
    \\&=
    (a; q)_\infty
    \sum_{k \geq 0} \sum_{m \geq 0}
    [x^m] \frac{1}{(a x; q)_\infty} q^{m k} \frac{z^k}{k!}
    \\&=
    (a; q)_\infty
    \sum_{m \geq 0}
    a^m [x^m] \frac{1}{(x; q)_\infty} e^{q^m z}
\end{align*}
Applying Lemma~\ref{th:q_identities}
we conclude
\[
    \poch(z,a,q) =
    (a; q)_\infty
    \sum_{m \geq 0}
    \frac{a^m}{(q; q)_m} e^{q^m z}
\]
Injecting this in the expression of $P_n(q)$, we obtain
\begin{align*}
    P_n(q) &=
    n! [z^n]
    \poch \left( - \frac{q}{1-q} z, \frac{1-q}{q}, q\right)
    e^{\frac{q}{1-q} z}
    \\&=
    \left( \frac{q}{1-q} \right)^n
    \left( \frac{1-q}{q}; q \right)_\infty
    n! [z^n]
    \sum_{m \geq 0}
    \frac{\left( \frac{1-q}{q} \right)^m}{(q;q)_m}
    e^{- q^m z}
    e^z
    \\&=
    \left( \frac{q}{1-q} \right)^n
    \left( \frac{1-q}{q}; q \right)_\infty
    \sum_{m \geq 0}
    \frac{\left( \frac{1-q}{q} \right)^m}{(q;q)_m}
    (1-q^m)^n
    \\&=
    q^n
    \left( \frac{1-q}{q}; q \right)_\infty
    \sum_{m \geq 0}
    \frac{\left( \frac{1-q}{q} \right)^m}{(q;q)_m}
    [m]_q^n.
\end{align*}
To conclude, Lemma~\ref{th:q_identities} is applied
\[
    P_n(q) =
    \frac{1}{e_q(1/q)}
    \sum_{m \geq 0}
    \frac{[m]_q^n}{[m]_q!} q^{n-m}.
\]
We apply d'Alembert's criteria
to find the values of $q$ for which this formal sum converges:
\[
    \lim_{m \to \infty}
    \frac{\frac{[m+1]_q^n}{[m+1]_q!} q^{n-m-1}}
    {\frac{[m]_q^n}{[m]_q!} q^{n-m}}
    = \frac{1-q}{q}
\]
is smaller than $1$ when $q > 1/2$.
\end{proof}

%%%%%%%%%%%%
    \paragraph{Limit law.}
%%%%%%%%%%%%

\begin{lemma} \label{th:q_asymptotics}
As $n$ and $m$ tend to infinity, $q = e^s$, $m s$ and $n s$ tend to $0$,
we have
\begin{align*}
    [m]_q^n &=
    m^n \exp \left(
    \frac{1}{2} n m s
    + \frac{1}{12} n m^2 \frac{s^2}{2} \right)
    \left( 1 + \bigO(n m^3 s^3 + n s) \right)
    \\
    [m]_q! &=
    m^m e^{-m} \sqrt{2 \pi [m]_q}
    \exp \left(
    \frac{1}{4} m^2 s
    + \frac{1}{36} m^3 \frac{s^2}{2} \right)
    \left( 1 + \bigO(m^4 s^3 + m s) + \smallo(1)\right).
\end{align*}
\end{lemma}

\begin{proof}
Let $S(x)$ denote the function $(e^x - 1 - x) / x$, then
\[
    [m]_q^n =
    \left(\frac{1 - e^{s m}}{1 - e^{s}}\right)^n
    =
    m^n \left(\frac{1 + S(s m)}{1 + S(s)} \right)^n
    =
    m^n e^{n \log(1 + S(sm)) - n \log(1 + S(s))}.
\]
We use the development
\[
    \log(1 + S(x)) =
    \log \left( 1 + \frac{x}{2} + \frac{x^2}{6} + \bigO(x^3) \right)
    =
    \frac{x}{2} + \frac{x^2}{24} + \bigO(x^3)
\]
to obtain
\begin{align*}
    [m]_q^n &=
    m^n \exp \left(
    \frac{1}{2} n m s
    + \frac{1}{12} n m^2 \frac{s^2}{2}
    + \bigO(n m^3 s^3 + n s) \right)
\end{align*}
According to Moak~\cite{moak1984q}, we have the following $q$-analog
of Stirling formula when $x \to \infty$ while $x \log(q) \to 0$
\[
    \log(\Gamma_q(x)) =
    (x - 1/2) \log([x]_q)
    + \frac{\Li_2(1 - q^x)}{\log(q)}
    + \frac{1}{2} \log(2 \pi)
    + \smallo(1)
\]
where $\Li_2(z)$ denotes the Dilogarithm function
\[
    \Li_2(z) = \sum_{k \geq 1} \frac{z^k}{k^2}
\]
We deduce
\begin{align*}
    [m]_q! &=
    [m]_q \Gamma_q(m)
    =
    [m]_q
    \exp \left(
    (m - 1/2) \log([m]_q)
    + \frac{\Li_2(1 - q^m)}{\log(q)}
    + \frac{1}{2} \log(2 \pi) + \smallo(1)
    \right)
    \\&=
    [m]_q^m
    \exp \left( \frac{\Li_2(1 - q^m)}{\log(q)} \right)
    \sqrt{2 \pi [m]_q}
    (1 + \smallo(1))
\end{align*}
The first part of the lemma provides
\[
    [m]_q^m =
    m^m \exp \left(
    \frac{1}{2} m^2 s
    + \frac{1}{12} m^3 \frac{s^2}{2} \right)
    \left( 1 + \bigO(m^4 s^3 + m s) \right).
\]
The Dilogarithm is expanded as
\[
    \frac{\Li_2(1-q^m)}{\log(q)}
    =
    \frac{1}{s}
    \sum_{k \geq 1}
    \frac{1}{k^2}
    \left( - m\, s \left( 1 + S(m\, s) \right) \right)^k
    =
    - m
    \left(
    1
    + \frac{1}{4} m s
    + \frac{1}{18} m^2 \frac{s^2}{2}
    + \bigO(m\, s)^3
    \right).
\]
Injecting those past two expansions
in the previous one concludes the proof.
\end{proof}

\begin{theorem}
The asymptotic mean $E_n$ and standard deviation $\sigma_n$
of the number $X_n$ of queries
used by the universal \ac algorithm on a partition of size $n$
chosen uniformly at random are
\[
    E_n = \frac{1}{4} (2 \zeta - 1) e^{2 \zeta}
    \quad \text{and} \quad
    \sigma_n = \frac{1}{3} \sqrt{\frac{3 \zeta^2 - 4 \zeta + 2}{\zeta + 1} e^{3 \zeta}},
\]
where $\zeta$ is the unique positive solution of
\[
    \zeta e^{\zeta} = n.
\]
The normalized random variable
\[
    X_n^\star = \frac{X_n - E_n}{\sigma_n}
\]
follows in the limit a normalized Gaussian law.
\end{theorem}

\begin{figure}
\begin{center}
\includegraphics[width=0.8\linewidth]{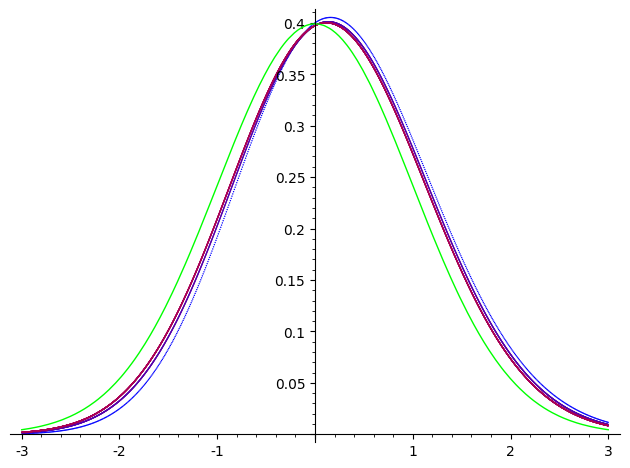}
\end{center}
\caption{In green, the probability density function of the normal distribution. In blue, purple and red, the empirical density functions for the number of queries, normalized by their mean and standard deviation, for $n$ in $\{100, 300, 600\}$.
We observe a slow convergence to the Gaussian limit law.}
\label{fig:limit_law}
\end{figure}

\begin{figure}
\begin{center}
\includegraphics[width=0.8\linewidth]{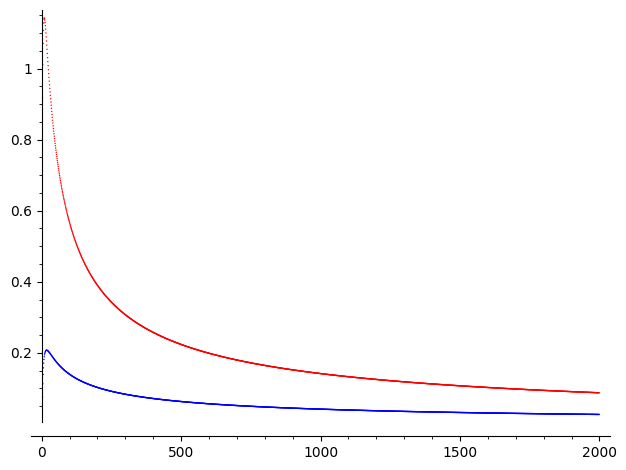}
\end{center}
\caption{A plot testing the asymptotic mean and standard deviations stated by Theorem~\ref{th:chordal_limit_law}. Let $\alpha_n$ and $\beta_n$ denote the sequences defined by
$\mean(X_n) = \frac{1}{4} (2 W(n) - 1 + \alpha_n) e^{2 W(n)}$
and $\sigma(X_n) = \frac{1}{3} \sqrt{\frac{3 W(n)^2 - 4 W(n) + 2 - \beta_n}{W(n) + 1} e^{3 W(n)}}$.
Plot of $\alpha_n$ in blue, $\beta_n$ in red,
for $n$ from $3$ to $2000$.
We observe a slow convergence to $0$.}
\label{fig:mean_variance}
\end{figure}

Those results are tested numerically in Figures~\ref{fig:limit_law}
and~\ref{fig:mean_variance}.

\begin{proof}
To prove the limit law, we show that the Laplace transform $\mean(e^{s X_n^\star})$
converges pointwise to the Laplace transform of the normalized Gaussian $e^{s^2/2}$.
We have
\[
    \mean(e^{s X_n^\star})
    =
    e^{- s E_n / \sigma_n}
    \mean(e^{s X_n / \sigma_n})
    =
    \frac{e^{- s E_n / \sigma_n}}{B_n}
    P_n(e^{s / \sigma_n})
\]
For any fixed real value $s$,
we compute the asymptotics of $P_n(e^{s/\sigma_n})$.
Let $q := e^{s/\sigma_n}$, so $q$ tends to $1$, then
\[
    P_n(e^{s/\sigma_n}) =
    \frac{1}{e_q(e^{-s/\sigma_n})}
    \sum_{m \geq 0}
    \frac{[m]_q^n}{[m]_q!}
    e^{(n - m) s / \sigma_n}.
\]
Motivated by the asymptotics from Lemma~\ref{th:q_asymptotics},
we rewrite this expression as
\begin{equation}
\label{eq:Pnq_A_phi}
    P_n(e^{s/\sigma_n}) =
    \frac{1}{e_q(e^{-s/\sigma_n})}
    \sum_{m \geq 0}
    A_{n,s}(m)
    e^{-\phi_{n,s}(m)}
\end{equation}
where
\begin{align*}
    A_{n,s}(m) &=
    \frac{[m]_q^n}{m^n \exp \left( \frac{1}{2} n m \frac{s}{\sigma_n} + \frac{1}{12} n m^2 \frac{(s/\sigma_n)^2}{2} \right)}
    \frac{m^m e^{-m} \exp \left( \frac{1}{4} m^2 \frac{s}{\sigma_n} + \frac{1}{36} m^3 \frac{(s/\sigma_n)^2}{2}\right)}{[m]_q!}
    e^{(n - m) s / \sigma_n},
    \\
    \phi_{n,s}(m) &=
    - n \log(m) + m \log(m) - m
    - \frac{1}{4} (2n - m) m \frac{s}{\sigma_n}
    - \frac{1}{36} (3n - m) m^2 \frac{(s/\sigma_n)^2}{2}.
\end{align*}
The dominant contribution to the sum comes from integers $m$
close to the minimum of $\phi_{n,s}(m)$, so we study this function.
The successive derivatives of $\phi_{n,s}(m)$ are
\begin{align*}
    \phi_{n,s}'(m) &=
    - \frac{n}{m} + \log(m) - \frac{1}{2} (n-m) \frac{s}{\sigma_n} - \frac{1}{12} (2n-m) m \frac{(s/\sigma_n)^2}{2},
    \\
    \phi_{n,s}''(m) &=
    \frac{n}{m^2} + \frac{1}{m} + \frac{s}{2 \sigma_n} - \frac{1}{6} (n-m) \frac{(s/\sigma_n)^2}{2}
    \\
    \phi_{n,s}'''(m) &=
    - \frac{2n}{m^3} - \frac{1}{m^2} + \frac{(s/\sigma_n)^2}{12}
\end{align*}
When $n$ is large enough, the second derivative of $\phi_{n,s}(m)$
is strictly positive for all $m > 0$, so the function is convexe.
It reaches its unique minimum at a value denoted by $m(s)$
and characterized by $\phi_{n,s}'(m(s)) = 0$.
Injecting the Taylor expansion
\[
    m(s) =
    m_0 + m_1 \frac{s}{\sigma_n} + m_2 \frac{(s/\sigma_n)^2}{2}
    + \cdots
    %\bigO \left( e^{4 \zeta} (s / \sigma_n)^3 \right)
\]
in this equation, rewriting $n$ as $\zeta e^{\zeta}$
and extracting the coefficients of the powers of $s$,
we obtain
\[
    m_0 = e^{\zeta},
    \qquad
    m_1 = \frac{1}{2} \frac{\zeta - 1}{\zeta + 1} e^{2 \zeta},
    \qquad
    m_2 =
    \frac{1}{3}
    \frac{2 \zeta^3 - 3 \zeta^2 + 2}{(\zeta + 1)^3}
    e^{3 \zeta}
\]
The dominant contribution to the sum defining $P_n(e^{s/\sigma_n})$
comes from values $m$ close to $m(s)$.
The \emph{central part} $C_n$ is defined as
the integers $m$ such that $|m - m(s)| < c_n$.
A heuristic proposed by~\cite{FS09} is to find $c_n$ such that
\[
    |\phi_{n,s}''(m(s))| c_n^2 \to +\infty
    \quad \text{and} \quad
    |\phi_{n,s}'''(m(s))| c_n^3 \to 0.
\]
As $n$, and thus $\zeta$, tend to infinity, we have
\[
    m(s) \sim e^{\zeta} \sim \frac{n}{\log(n)},
    \qquad
    |\phi_{n,s}''(m(s))| \sim \zeta e^{-\zeta} \sim \frac{\log(n)^2}{n},
    \qquad
    |\phi_{n,s}'''(m(s))| \sim \zeta e^{-2 \zeta} \sim \frac{\log(n)^3}{n^2},
\]
so we choose $c_n = e^{3 \zeta / 5}$.
Uniformly for $m$ in $C_n$, we have
\begin{align*}
    m(s) &=
    e^{\zeta}
    + \frac{1}{2} \frac{\zeta - 1}{\zeta + 1} e^{2 \zeta} \frac{s}{\sigma_n}
    + \frac{1}{3}
    \frac{2 \zeta^3 - 3 \zeta^2 + 2}{(\zeta + 1)^3}
    e^{3 \zeta}
    \frac{(s / \sigma_n)^2}{2}
    + \bigO(e^{-\zeta/2}),
    \\
    \phi_{n,s}(m) &=
    - (\zeta^2 - \zeta + 1) e^{\zeta}
    - E_n \frac{s}{\sigma_n}
    - \frac{s^2}{2}
    + (\zeta + 1) e^{-\zeta} \frac{(m - m(s))^2}{2}
    + \bigO(e^{-\zeta / 4}),
    \\
    A_{n,s}(m) &=
    \frac{1}{\sqrt{2 \pi e^{\zeta}}}
    \left( 1 + \smallo(1) \right).
\end{align*}
In fact, computing the Taylor expansion of $\phi_{n,s}(m)$ at $m(s)$ came first.
We chose the values of $E_n$ and $\sigma_n$
so that the coefficients in $s$ and $s^2$
are the ones presented in the above equation.
The error term is then obtained using the Lagrange form of the remainder
in Taylor's Theorem.
We deduce the following asymptotics for the central part of the sum
\[
    \sum_{m \in C_n}
    A_{n,s}(m) e^{-\phi_{n,s}(m)}
    \sim
    \frac{1}{\sqrt{2 \pi e^{\zeta}}}
    e^{(\zeta^2 - \zeta + 1) e^{\zeta}
        + E_n \frac{s}{\sigma_n}
        + \frac{s^2}{2}}
    \sum_{m \in C_n}
    e^{- (\zeta + 1) e^{-\zeta} \frac{(m - m(s))^2}{2}}.
\]
Applying the Euler-Maclaurin formula to turn the sum into an integral,
we obtain
\[
    \sum_{m \in C_n}
    A_{n,s}(m) e^{-\phi_{n,s}(m)}
    \sim
    \frac{1}{\sqrt{2 \pi e^{\zeta}}}
    e^{(\zeta^2 - \zeta + 1) e^{\zeta}
        + E_n s / \sigma_n
        + s^2 / 2}
    \int_{-c_n}^{c_n}
    e^{- (\zeta + 1) e^{-\zeta} \frac{x^2}{2}}
    d x.
\]
After the variable change $y = \sqrt{(\zeta + 1) e^{-\zeta}} x$,
observing that $\sqrt{(\zeta + 1) e^{-\zeta}} c_n$ tends to infinity,
the integral is approximated as a Gaussian integral and we conclude
\[
    \sum_{m \in C_n}
    A_{n,s}(m) e^{-\phi_{n,s}(m)}
    \sim
    \frac{1}{\sqrt{\zeta + 1}}
    e^{(\zeta^2 - \zeta + 1) e^{\zeta}
        + E_n s / \sigma_n
        + s^2 / 2}.
\]
When we compare the asymptotics of the central part
to the asymptotics of the Bell numbers (see, \eg, \cite{FS09})
\[
    B_n \sim
    \frac{e^{(\zeta^2 - \zeta + 1) e^{\zeta} - 1}}{\sqrt{\zeta + 1}},
\]
we see from Equation~\eqref{eq:Pnq_A_phi} that, as expected,
\[
    \frac{e^{- s E_n / \sigma_n}}{B_n}
    \frac{1}{e_q(e^{-s/\sigma_n})}
    \sum_{m \in C_n}
    A_{n,s}(m) e^{-\phi_{n,s}(m)}
    \sim
    e^{s^2/2}.
\]
Let us now prove that the part of the sum
corresponding to $m \leq m(s) - c_n$
is negligible compared to the central part.
According to Lemma~\ref{th:q_asymptotics}, we have
\[
    A_{n,s}(m) =
    \left(1 + \bigO(n m^3 / \sigma_n^3) \right)
    \left(1 + \bigO(m^4 / \sigma_n^3) \right)
    \frac{1}{\sqrt{2 \pi [m]_q}}
    = \bigO(m^8).
\]
Since $\phi_{n,s}(m)$ is convexe (for large enough $n$),
we have $\phi_{n,s}(m) \geq \phi_{n,s}(m(s) - c_n)$
for all $m < m(s) - c_n$.
Since
\[
    \phi_{n,s}(m(s) - c_n) =
    - (\zeta^2 - \zeta + 1) e^{\zeta}
    - E_n \frac{s}{\sigma_n}
    - \frac{s^2}{2}
    + (\zeta + 1) e^{-\zeta} \frac{c_n^2}{2}
    + \bigO(e^{-\zeta / 4})
\]
and $e^{-\zeta} c_n^2$ tends to infinity as $e^{\zeta / 5}$,
we obtain for all $m < m(s) - c_n$
\[
    \phi_{n,s}(m) \geq
    - (\zeta^2 - \zeta + 1) e^{\zeta}
    - E_n \frac{s}{\sigma_n}
    - \frac{s^2}{2}
    + \exactbigO(e^{\zeta / 5}).
\]
We conclude
\begin{align*}
    \sum_{m < m(s) - c_n} A_{n,s}(m) e^{-\phi_{n,s}(m)}
    &\leq
    \sum_{m < m(s) - c_n} \bigO(m^8)
    e^{(\zeta^2 - \zeta + 1) e^{\zeta}
    + E_n s / \sigma_n
    + s^2/2
    - \exactbigO(\exp(\zeta / 5}))
    \\&\leq
    e^{(\zeta^2 - \zeta + 1) e^{\zeta}
    + E_n s / \sigma_n
    + s^2/2}
    m(s)^9
    e^{- \exactbigO(\exp(\zeta / 5))}.
\end{align*}
Since $m(s) \sim e^{\zeta}$, this result
is exponentially small, with respect to $n$, compared to the central part.
Let us now prove that the part of the sum beyond the central part
is negligible as well.
There is a constant $C$ such that for $n$ large enough
and any $m \geq C e^{3 \zeta /2}$, we have
\[
    - \frac{1}{4} (2n-m) m \frac{s}{\sigma_n}
    - \frac{1}{36} (3n-m) m^2 \frac{(s / \sigma_n)^2}{2} \geq 0.
\]
In that case, we obtain the simple bound
\[
    \phi_{n,s}(m) \geq - n \log(m) + m \log(m) - m \geq m - n \log(m).
\]
Injecting this bound and $A_{n,s}(m) = \bigO(m^8)$ in the sum, we obtain
\[
    \sum_{m \geq C e^{3 \zeta / 2}} A_{n,s}(m) e^{-\phi_{n,s}(m)}
    \leq
    \bigO(1)
    \sum_{m \geq C e^{3 \zeta / 2}} m^{n+8} e^{-m}.
\]
The sum is bounded by an integral and $n+8$ integration by part are applied
\begin{align*}
    \sum_{m \geq C e^{3 \zeta / 2}} A_{n,s}(m) e^{-\phi_{n,s}(m)}
    &\leq
    \bigO(1)
    (n+8)!
    (C e^{3 \zeta / 2})^{n+8}
    e^{- C \exp(3 \zeta / 2)}
    \\&\leq
    \bigO(1)
    n^n
    (C e^{3 \zeta / 2})^{n+8}
    e^{- C \exp(3 \zeta / 2)}
\end{align*}
Since $n = \zeta e^{\zeta}$, we have $n^n = e^{\bigO(\zeta^2) \exp(\zeta)}$,
so
\[
    \sum_{m \geq C e^{3 \zeta / 2}} A_{n,s}(m) e^{-\phi_{n,s}(m)}
    \leq
    \bigO(1)
    e^{- \exactbigO(\exp(3 \zeta / 2))}
\]
which is negligible compared to the central part of the sum.
The last part we consider is $m(s) + c_n \leq m \leq C e^{3 \zeta /2}$.
Since $\phi_{n,s}(m)$ is decreasing there and $A_{n,s}(m) = \bigO(m^8)$,
we have
\[
    \sum_{m = m(s) + c_n}^{C e^{3 \zeta / 2}} A_{n,s}(m) e^{-\phi_{n,s}(m)}
    =
    \bigO(e^{27 \zeta / 2})
    e^{-\phi_{n,s}(m(s) + c_n)}
\]
As for the case $m = m(s) - c_n$, we find
\[
    \phi_{n,s}(m(s) + c_n) = - (\zeta^2 -\zeta + 1) e^{\zeta} - E_n \frac{s}{\sigma_n} - \frac{s^2}{2} + \exactbigO(e^{\zeta/5})
\]
and conclude
\[
    \sum_{m = m(s) + c_n}^{C e^{3 \zeta / 2}} A_{n,s}(m) e^{-\phi_{n,s}(m)}
    =
    e^{(\zeta^2 -\zeta + 1) e^{\zeta} + E_n \frac{s}{\sigma_n} + \frac{s^2}{2}}
    \bigO(e^{27 \zeta / 2})
    e^{- \exactbigO(e^{\zeta/5})},
\]
which is negligible compared to the central part.
In conclusion, we have
\[
    \mean(e^{s X_n^\star})
    =
    \frac{e^{- s E_n / \sigma_n}}{B_n}
    P_n(e^{s / \sigma_n})
    \sim
    \sum_{m \in C_n}
    A_{n,s}(m) e^{-\phi_{n,s}(m)}
    \sim
    e^{s^2/2}.
\]
Since the Laplace transform of $X_n^\star$ converges pointwise
to the Laplace transform of the normalized Gaussian law,
$X_n^\star$ converges in distribution to this Gaussian.
\end{proof}

%%%%%%%%%%%%%%%%%%%%%%%%
%%%%%%%%%%%%%%%%%%%%%%%%
%%%%%%%%%%%%%%%%%%%%%%%%
\subsection{Proof of Theorem~\ref{clique_algo_complexity}}
Suppose first that $p_1>\cdots>p_k$. At time $t$, we have identified classes $C_1^t,\dots,C_k^t$, say with sizes $c+i^t=|C_i^t$.  It takes time at most $k\sqrt n$ to process the first $\sqrt n$ vertices, and at that point and all subsequent times we have (with exponentially small failure probability) that $c_i^t\sim p_i\sqrt n$ for each $i$, where $C_i^t$ consists of elements of type $i$.  In particular $|C_1^t|>|C_2^t|>\cdots>|C_k^t|$.
The expected number of comparisons used for each subsequent element is $\sum{i=1}^kp_i$, as the probability that the element is of type $i$ is $p_i$, and in this case we need $i$ comparisons to place it in $C_i^t$.

The case where some of the $p_i$ are the same is similar, except that if $p_i=p_j$ then $c_i^t$ and $c_j^t$ may switch order over time.  However this is fine: we need only have the property that if $p_i>p_j$ then $c_i^t>c_j^t$, and the same argument then works, possibly permuting colors for classes $i,j$ with $p_i=p_j$.

\subsection{Proof of Theorem~\ref{random_algo_complexity_crossedges}}

The idea of the proof is as follows: we estimate the behaviour of the process until large linear time, and then note that it finishes in negligible time.  Let us note first that in any partition of a set of size $t$, where $t$ is large, a fraction of at least (about) $1/k$ of the pairs lie inside classes.  It follows that at each time step, we reduce the number of vertices by 1 with constant probability, and so the the algorithm finishes in expected time $O(kt)$.  When handling an instance of size $n$, it is therefore enough to run the algorithm until $o(n)$ vertices remain, and then note that expected time to complete is still $o(n)$.

We consider an instance of size $n$, and analyze the following process.  Begin with all vertices marked {\em active}.  At each time step, pick (with replacement) a random pair $\{u,v\}$ and:
\begin{itemize}
    \item If $u,v$ are active and from distinct classes $i,j$ then say we have generated an {\em $ij$-crossedge}.
    \item If $u,v,$ are both active and in class $i$ then mark exactly one of $u$ and $v$ inactive.
    \item If one of $u,v$ is inactive, then do nothing.
\end{itemize}
Note that as the process runs, the number of active vertices is monotonic decreasing, and we are increasingly likely to choose pairs where one vertex is inactive.  These contribute to the new process, but do not generate new comparisons between pairs. So we are looking at a (randomly) slowed down version of the random algorithm; but this makes the analysis much simpler!

Let $x_i(t)$ denote the number of active class $i$ vertices after $t$ time steps, and $x_{ij}(t)$ denote the number of $ij$-crossedges that are generated in the first $t$ time steps.  Then at step $t+1$, the probability that we pick two active vertices in class $i$ is
$$\binom{x_i(t)}{2}\Big/\binom n2 \sim \frac{x_i(t)^2}{n^2}.$$
Writing $p=p_i$, we estimate $x_i(t)$ via a function $x=x(t)$ satisfying the differential equation
\begin{align*}
x(0)   &= pn\\
x'(t) &= -\frac{x(t)^2}{n^2}.
\end{align*}
This has solution
$$x(t)=\frac{n^2}{t+n/p}=\frac{pn}{1+pt/n}.$$
Note that at time $\lambda n$, as $\lambda$ gets large, we get
$$x(\lambda n)\sim\frac{pn}{1+\lambda p}.$$
The actual value of $x_i(t)$ closely tracks the differential equation with very high probability.  This follows straightforwardly from a standard application of the R\"odl nibble or differential equation method (see for example \cite{alonspencer}): we throw in the edges in batches of size $\epsilon n$, and note that a Chernoff bound implies that we stick close to the the differential equation as we are in total making $O(n)$ comparisons.

Now let's estimate the number of $ij$-crossedges.  Using our estimates for $x_i(t)$ and $x_j(t)$, we see that the probability of an $ij$ crossedge at step $t+1$ is 
$$
x_i(t)x_j(t)/\binom n2
\sim \frac{2x_i(t)x_j(t)}{n^2}
\approx \frac{2n^2}{(t+n/p_i)(t+n/p_j)}.
$$
Let $p=p_i$ and $q=p_j$.   We can model the growth of $x_{ij}(t)$ by a function $c=c(t)$ satisfying the differential equation
\begin{align*}
    c(0)&= 0\\
    c'(t)&=\frac{2n^2}{(t+n/p)(t+n/q)}.
\end{align*}

If $p=q$, we have
$$c'(t)=\frac{2n^2}{(t+n/p)^2}$$
and 
\begin{align*}
    c(t)=
    \int_0^t c'(s)ds=\int_0^t \frac{2n^2}{(s+n/p)^2}ds
    =\left[\frac{-2n^2}{s+n/p}\right]_0^t,
\end{align*}
giving
\begin{align*}
    c(t)=2pn\left(1-\frac{1}{2+2pt/n}\right)\sim 2pn
\end{align*}
as $t/n\to \infty$.

If $p\ne q$, we have 
\begin{align*}
    c(t)
    &=\int_0^t \frac{2n^2}{(s+n/p)(s+n/q)}ds\\
    &=\frac{2npq}{p-q}\int_0^t \frac{1}{s+n/p}-\frac{1}{s+n/q}ds\\
    &=\frac{2npq}{p-q}\ln\left(\frac{1+pt/n}{1+qt/n}\right)\\
    &\sim \frac{2npq\ln(p/q)}{p-q}
\end{align*}
as $t/n\to\infty$.
Combining this with the remarks at the beginning of the proof gives the result.

%%%%%%%%%%%%%%%%%%%%%%%%%%%%%%%%%%%%%%%%%%%%%%%%%%%%%%
    \subsection{Proofs of Section~\ref{sec:noisy}}
%%%%%%%%%%%%%%%%%%%%%%%%%%%%%%%%%%%%%%%%%%%%%%%%%%%%%%

Consider an \ac algorithm applied to a set of size $n$.
Let $Q$ denote the set of queries with their answers
when the algorithm terminates.
We associate to $Q$ a graph $G$
where each vertex corresponds to an element of the starting set,
each edge is either \emph{positive} or \emph{negative}
and corresponds to the answers from $Q$.
As stated in Paragraph \emph{Correcting errors},
a contradiction is detected in $Q$
if and only if $G$ contains a contradictory cycle
(a cycle where all edges except one are positive).
In that case, additional queries are submitted
until the responsible answers are identified and corrected.

If $Q$ contains no more contradiction,
it characterizes a set partition $p$
(otherwise, the \ac algorithm has not terminated)
of size $n$ with a number of blocks denoted by $b$.
However, this partition might not be the correct solution
if $Q$ contains undetected errors.
Assume each answer is wrong with a small probability $p$
and let $\probaerror$ denote the probability that
$Q$ contains errors but no contradictory cycle.
Its Taylor coefficients at $p=0$ are denoted by $(c_i)_{i \geq 0}$
\[
    \probaerror = c_0  + c_1 p + c_2 p^2 + \cdots
\]
We would like $Q$ to contain few additional queries
compared to the noiseless case,
while ensuring that $\probaerror$ is small.
Since $p$ is assumed to be small,
the second conditions corresponds to
minimizing the vector $(c_0, c_1, \ldots)$
for the lexicographic order.
Indeed, if we consider a larger vector $(c_0', c_1', \ldots)$,
then there is a positive $\epsilon$ such that
for any $0 < p < \epsilon$, we have
\[
    c_0 + c_1 p + c_2 p^2 + \cdots <
    c_0' + c_1' p + c_2 p^2 + \cdots
\]
Let $d_k$ denote the number of sets of $k$ edges from $G$
which could be switched (positive edges become negative ones,
negative edges become positive ones)
without creating a contradictory cycle.
This corresponds to the number of ways to change $k$ answers in $Q$
and obtain a result without contradiction.
Let $m$ denote the total number of edges in $G$.
Then the probability that there is an undetected error is
\[
    \sum_{k \geq 0}
    d_k p^k (1 - p)^{m-k}.
\]
Since this probability is $c_0 + c_1 p + c_2 p^2 + \cdots$,
we deduce for all $k \geq 0$
\[
    c_k =
    \sum_{j = 0}^k
    d_j [x^k] x^j (1-x)^{m-j}.
\]
The triangular shape of this system of equations
implies that the vector $(c_0, c_1, \ldots)$
is minimal for the lexicographic order
if and only if the vector $(d_0, d_1, \ldots)$
is minimal for the lexicographic order.

Since $G$ contains no contradictory cycle,
we have $d_0 = 0$.
If all positive components of $G$ not reduced to one vertex are $2$-edge-connected
(\ie they have no vertex of degree $1$),
then switching a positive answer must create a contradictory cycle.
If every pair of positive components from $G$
is linked by at least $2$ negative edges,
then switching a negative answer must create a contradictory cycle.
Thus, if $G$ satisfies those two conditions, we have $d_1 = 0$.

In order to make $d_2$ vanish, we would need every positive component of $G$
to be $3$-edge-connected, so to have minimal degree at least $3$
(unless the component is reduced to one vertex).
As we saw in Paragraph \emph{Bounded number of errors}
from Section~\ref{sec:noisy},
this would substantially increase the number of queries.
Thus, we choose a different approach, fixing a parameter $r$
that influences the number of queries added compared to the noiseless case,
then constructing $Q$ so that $d_2$ (and hence $c_2$) is minimized.
We will first describe the structure of $G$ (and hence $Q$),
then provide an algorithm reaching this structure.

First, we ensure that each pair of positive components of $G$
are linked by at least $3$ negative answers.
This adds at most $2 \binom{b}{2}$ queries,
but should in practice be small,
as in our random models,
in the noiseless case,
all pairs of positive components are typically already linked
by more than $3$ negative edges.
This ensures that there are no pairs of negative answers from $Q$
that can be switched to create a contradictory cycle.
Thus, $d_2$ counts the number of pairs of positive edges
that can be switched without creating a contradictory cycle.
Let $n$ denote the number of vertices of $G$
and $m$ its number of positive edges.
If the partition $p$ characterized by $G$ contains $b$ blocks,
then the minimal possible value of $m$ is $n-b$.
It corresponds to all positive components being trees
and is reached in the noiseless case.
Since we want $m$ to not grow too far from this lower bound,
we assume $m < 3n/2$.
According to~\cite[Section 4]{bauer1985combinatorial},
the structure of $G$ minimizing $d_2$ is
a graph where all vertices have degree $2$ or $3$,
and there is an integer $s$ such that
the maximal paths or cycles containing only vertices of degree $2$
all have length $s$ or $s - 1$.
In practice, during the query of human experts,
this constraint might be difficult to satisfy exactly.
So let us denote by $r$ the average length of those $2$-paths,
and by $r'$ an upper-bound.
The number of positive edges is then
\[
    m = n \left(1 + \frac{1}{3 r + 2} \right),
\]
so the number of edges added compared to the noiseless case is
\[
    m - (n - b) = \frac{n}{3 r + 2} + b.
\]
The number of $2$-paths is
\[
    \frac{3 n}{3 r + 2},
\]
so a bound on $d_2$ is
\[
    \binom{r' + 1}{2} \frac{3 n}{3 r + 2}.
\]

We now describe an algorithm reaching this desirable structure.
It inputs a positive parameter $r$.
The largest $r$ is, the smaller the number of additional edges is
compared to the noiseless case,
but also the higher the probability of undetected error becomes.
We start with an \ac algorithm designed for the noiseless case
(the clique algorithm for example) and maintain the graph $G$ described above.
In the first phase, the positive components of $G$ are trees.
Whenever the noiseless algorithm proposes a query
between two positive components of $G$,
we choose one vertex on each of those components such that
a positive answer would create a positive component
that is a tree where all vertices have degree at most $3$,
and all $2$-paths have length close to $r$.
The second phase starts when the noiseless algorithm has terminated.
Additional queries are added
\begin{itemize}
\item
between positive components so that
each pair is linked by at least $3$ negative answers,
\item
between the leaves of each tree corresponding to a positive component of $G$.
No vertex should be left with positive degree $1$,
the $2$-paths should have length close to $r$,
and each positive component should be $2$-edge-connected.
\end{itemize}
If at any point a contradictory cycle is detected,
queries cutting it in two are submitted until the conflict is resolved.

\end{document}